\documentclass[lettersize,journal]{IEEEtran}
\usepackage{amsmath,amsfonts}
\usepackage{algorithmic}
\usepackage{algorithm}
\usepackage{array}
\usepackage[caption=false,font=normalsize,labelfont=sf,textfont=sf]{subfig}
\usepackage{textcomp}
\usepackage{stfloats}
\usepackage{url}
\usepackage{verbatim}
\usepackage{graphicx}
\usepackage{cite}
\usepackage{booktabs}
\usepackage{xcolor}
\usepackage{tcolorbox}

\usepackage{xcolor}
\newcommand{\zyf}[1]{{#1}}
\newcommand{\lqy}[1]{{#1}}
\newcommand{\qy}[1]{{#1}}
\newcommand{\minorqy}[1]{{#1}}
\usepackage{multirow} 
\usepackage{wrapfig}

\newcounter{finding}
\usepackage{mdframed}
\newcommand{\finding}[1]{\refstepcounter{finding}
  \vspace{0.5mm}
 \begin{mdframed}[linecolor=gray,roundcorner=12pt,backgroundcolor=gray!15,linewidth=3pt,innerleftmargin=2pt, leftmargin=0cm,rightmargin=0cm,topline=false,bottomline=false,rightline = false]
  \textbf{Answer to RQ\arabic{finding}:} #1
 \end{mdframed}
 \vspace{0.5mm}
}

\begin{document}

\title{Directional Diffusion-Style Code Editing Pre-training}

\author{Qingyuan Liang, Zeyu Sun*, Qihao Zhu, Junhao Hu, Yifan Zhao, Yizhou Chen, Mingxuan Zhu, Guoqing Wang, Lu Zhang*
        % <-this % stops a space
\thanks{Qingyuan Liang, Qihao Zhu, Junhao Hu, Yifan Zhao, Yizhou Chen, Mingxuan Zhu, Guoqing Wang, Lu Zhang are with
Key Lab of HCST (PKU), MOE; SCS, Peking University,
Beijing 100871, China (e-mail: liangqy@pku.edu.cn; zhuqh@pku.edu.cn;
junhaohu@stu.pku.edu.cn; zhaoyifan@stu.pku.edu.cn; yizhouchen@stu.pku.edu.cn; zhumingxuan@stu.pku.edu.cn; guoqingwang@stu.pku.edu.cn; zhanglucs@pku.edu.cn).}% <-this % stops a space
\thanks{Zeyu Sun is with National Key Laboratory of Space Integrated Information System, Institute of Software, Chinese Academy of Sciences,
Beijing, China (e-mail: zeyu.zys@gmail.com).}
\thanks{
% Lu Zhang is the corresponding author. 
We have made the DivoT5 model and all related resources publicly available on GitHub~(https://github.com/LIANGQINGYUAN/DivoT5)}
\thanks{*Corresponding authors: Zeyu Sun and Lu Zhang}
}



\maketitle

\begin{abstract}
Code pre-trained models have shown promising effectiveness in various software engineering tasks. Among these tasks, many tasks are related to software evolution and/or code editing. However, existing code pre-trained models often overlook the real-world code editing data and the evolutionary nature of the editing process.
In this paper, to simulate the step-by-step code editing process of human developers, we propose DivoT5, a pre-trained model based on directional diffusion at the data level. 
In DivoT5, we adopt two categories of pre-training tasks. The first category is mask and denoising tasks augmented with a diffusion direction representing code evolution. That is, we first apply a noising process to the code snippets before evolution, and then ask the pre-training process to restore the snippets with noise into the code snippets after evolution. 
The second category is tasks aiming to reinforce the evolutionary direction. That is, we first generate various intermediate versions for each pair of snippets before and after evolution, 
and then ask the pre-training process to transform the intermediate versions into the snippet after evolution for each pair.
We evaluate DivoT5 for two code-editing scenarios (including a number of tasks) and one non-editing scenario using four downstream tasks. 
% Given each downstream task, we fine-tune the pre-trained DivoT5 on the various datasets for the downstream task and evaluate the effectiveness of DivoT5. 
For each downstream task, we fine-tune the pre-trained DivoT5 on multiple corresponding datasets and evaluate its effectiveness across diverse scenarios
Our experimental results show that DivoT5 achieves state-of-the-art~(SOTA) performance on most tasks in comparison to models of the same scale (220M), large-scale (770M, 6.7B) models in fine-tuning, and billion-scale~(6.7B, 8B, ChatGPT) instruct models in few-shot settings.
For one code-editing task (i.e., CodeReview in NL-based CodeRefinement task), DivoT5 pre-trained on top of CodeT5-small (60M) can even outperform CodeT5-base (220M) and other pre-trained models with 220M parameters except for DivoT5 pre-trained on top of CodeT5-base (220M).
\end{abstract}


\section{Introduction}
\IEEEPARstart{T}{he} code of software systems often requires continuous maintenance and updates to enhance functionality, adapt to new requirements, and fix vulnerabilities~\cite{lientz1978softwaremainten,bennett2000softwareevolution,tufano2021codereview, recoder2021,just2014defects4j}, which are collectively termed ``code evolution''. Throughout the code evolution process, many tasks require generating new code by editing existing old code, such as code refinement~\cite{tufano2019coderefine,codereviewer, tufano2021codereview,coditt52022,twinxsql} and bug fixing~\cite{tufano2019bug2fix,tufano2019learning_bugfixing}. These tasks related to code editing constitute an essential part of the code evolution process. To address tasks related to code editing, many studies have utilized pre-trained code models, such as CodeT5~\cite{codet52021}, given their proven ability to effectively encode the semantic information of code. 
\qy{
However, directly applying these pre-trained models to code editing tasks often fails to achieve satisfactory performance because they do not adequately consider the contextual information~(i.e., the relationship between the old and new versions of code during the evolution process). 
This context includes the semantic alignment across code versions and serves as critical guidance for accurately identifying what should be modified, retained, or extended.
}

Recently, several studies~\cite{coditt52022,li2023codeeditor,codereviewer,lin2023cct5} propose pre-training tasks specifically designed for code editing tasks and learning contextual information of code evolution by fine-tuning pre-trained models.
These models consider code editing through two ways: masked language modeling~(MLM) with code changes~\cite{codereviewer,lin2023cct5} and denoising auto-encoding~(DAE) with controlled noises~\cite{li2023codeeditor,coditt52022}.
% MLM 
% 仅使用代码变更，
% 定义真实代码演化场景：旧代码+comment --> 新代码
% diff中不包含明确的意图（演化方向）
The first way represents code changes as the output sequence of the \textit{diff} tool and applies the MLM pre-training tasks to understand the semantics of code changes.
% modifications. 
% Specifically, the first way uses the output of the \textit{diff} tool, representing code changes as a sequence, and applies MLM pre-training tasks to understand the semantics of these code modifications.
% While this way shows promising results in understanding code changes, it still faces challenges in generating fully edited code. 
%%%%%% application scenarios
% Because the task of understanding code changes primarily involves using the code changes as input to generate corresponding descriptions~(e.g., commit messages~\cite{lin2023cct5}) or classifications~(e.g., whether the changes improve the code~\cite{codereviewer}), which does not align with the scenarios of editing existing code in real-world development.
% This way has shown promising results in understanding code changes. However, in real development scenarios, code evolution typically involves generating fully new code based on existing old code and the programmer's intent description.
\zyf{This way has shown promising results in understanding code changes. However, in real development scenarios, code evolution typically involves generating new code based on existing old code, rather than understanding the code change. Therefore, using code changes as input to generate corresponding descriptions~(e.g., commit messages~\cite{lin2023cct5}) or classification labels~(e.g., whether the code improves~\cite{codereviewer}) does not align well with such development scenarios for code generation.}
% However, in real development scenarios, code evolution typically involves generating new code based on existing code and the intended direction of evolution. 
% Therefore, using code changes as input to generate corresponding descriptions~(e.g., commit messages~\cite{lin2023cct5}) or classification labels~(e.g., whether the code improves~\cite{codereviewer}) does not align well with development scenarios.
\qy{
Moreover, the granularity they adopt is typically at the code line level~\cite{coeditor_iclr2024}, as produced by the \textit{diff} tool. However, capturing fine-grained changes at the token level is essential for accurately modeling developer intent and generating precise code modifications~\cite{edit_emperical,edit_emperical_2}. 
This also posing a challenge in considering fine-grained changes~(i.e., token-level) during the generation process~(i.e., \textbf{the first challenge} lies in reflecting fine-grained edits akin to those made by human developers).
}
% our：引入了token-level的diff
% DAE
\qy{
In the second way, noise is typically introduced to the old code based on the probability distribution derived from real code changes in downstream datasets, and the model employs the DAE-like pre-training tasks to restore the old code from the noisy input~\cite{coditt52022}. 
% While this noise-disrupted code merely reflects the overall probability of code being edited, without considering the varying contexts of the different code. 
However, this noise-disrupted code only captures the general likelihood of edits, without accounting for the fact that different pieces of old code should evolve into contextually appropriate new versions. In other words, the editing process is not merely about reflecting the overall probability of code being edited, but about transforming code in ways that reflect the specific intent and semantics embedded in each unique context.
Consequently, there exists a substantial gap that existing approaches often overlook between disrupted code and the imperfect, context-rich code written by real developers.
~(i.e., \textbf{the second challenge} lies in leveraging the context of code edits made by human developers).
}
% ours：真实编辑+丰富上下文
%%%%%% incremental editing 
In the development scenarios, code evolution typically involves a series of modifications to one file, making it a multi-step editing process rather than a one-off effort.
Instead of focusing on only \textit{what} the code changes are or \textit{what} the actual change probabilities might be, \textit{how} the code is gradually edited to achieve the final evolutionary goal should be also be considered. 
However, existing code editing pre-trained models only utilize the before-and-after code pairs directly, neglecting the gradual evolution of code in real-world scenarios and its characteristic of incremental editing
~(i.e., \textbf{the third challenge} lies in capturing the incremental nature of code editing performed by human developers.).
% and its corresponding rich context

\begin{figure*}[t]
  \centering
  % \vspace{-0.13cm}
  % \setlength{\abovecaptionskip}{10pt}
  \includegraphics[scale=0.9]{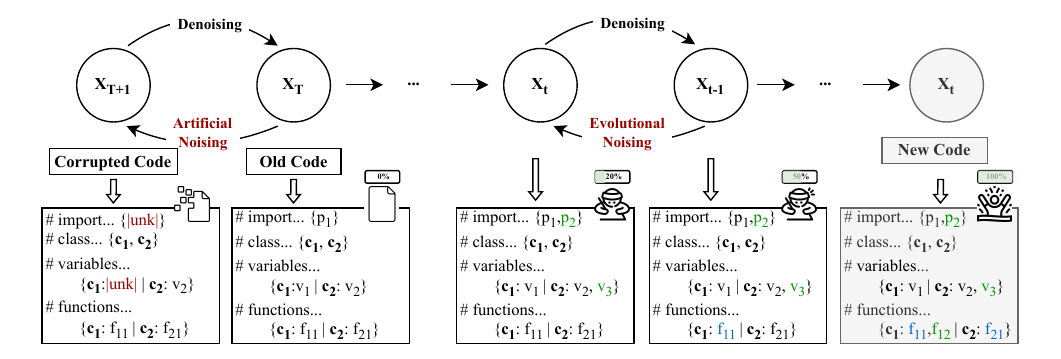}
  \caption{Illustration of directional diffusion. The upper part depicts the transition of code editing states, where noise in the old code is gradually reduced from left to right until the new code is obtained. Starting with the old code ($X_T$), artificial noise ($X_{T+1}$) is introduced via masking strategies to guide the evolutionary direction. Intermediate states ($X_t$) represent evolutionary noise, reinforcing the editing direction. The lower part simulates the changes in code content during human editing, with green indicating additions and blue representing modifications. From the old code to the new code, the programmer introduces the package \(p_2\), adds the variable \(v_3\) and the function \(f_{12}\), and updates the functions \(f_{11}\) and \(f_{21}\).}
  \label{overview_diffusion}
  % \vspace{-0.5cm}
\end{figure*}

% 1. 是生成编辑而不是理解编辑
% 2. 拿什么生成编辑
% 3. 如何生成
\lqy{
% In this paper, we introduce DivoT5, a diffusion-style pre-trained model designed to enhance code editing tasks in real-world scenarios, to address the above challenges. 
% In this paper, we present DivoT5, a diffusion-style pre-trained model designed to address the above challenges and enhance code editing tasks in real-world scenarios. 
In this paper, we introduce DivoT5, a diffusion-style pre-trained model. To tackle the above challenges, the key idea of DivoT5 is to simulate the incremental nature of human code editing in the pre-training step. 
It draws inspiration from diffusion models, which excel at handling incremental transformations in continuous data domains, generating high-quality and diverse images or audio through a gradual multi-step denoising process~\cite{diffapp_ho2020denoising, difflm_ARDiff_wu2024ar, difflm_DiffLM_li2022diffusion}.
% DivoT5 draws inspiration from diffusion models, which were originally developed for continuous data domains and are known for generating high-quality, diverse images or audio through a multi-step denoising process~\cite{diffapp_ho2020denoising, difflm_ARDiff_wu2024ar, difflm_DiffLM_li2022diffusion}. 
% diffapp_mittal2021symbolic, diffapp_nichol2021improved, diffapp_kong2020diffwave
% Similarly, code editing requires a series of modifications to meet the final requirements. 
% Similarly, code editing involves a series of iterative modifications to meet final requirements, making it well-suited to a diffusion-style approach.
% The directional diffusion style aims to further simulate the step-by-step code editing process of human developers, allowing the model to generate the desired new code in diverse contexts while understanding the characteristics of incremental code editing.
Similarly, code editing involves making a series of iterative changes to fulfill the final requirements,  
making it naturally aligned with the step-by-step refinement principles of diffusion models.
However, DivoT5 does not directly implement diffusion processes. It utilizes its data generation principles and focuses on simulating the gradual code editing steps that developers typically follow, allowing it to produce realistic and contextually relevant code modifications. To distinguish it from traditional diffusion methods, we refer to this approach as ``directional diffusion.''
}

Figure~\ref{overview_diffusion} illustrates the schematic of directional diffusion.
We set the starting point~(i.e., $X_T$) of the directional diffusion strategy as the old code, design three mask strategies as introducing artificial noise~(i.e., $X_{T+1}$) to understand the evolutionary direction, and treat the intermediate states~(i.e., $X_t$) of the editing process as evolutionary noise that reinforces the direction of evolution.
Specifically, $X_{T+1}$ to $X_{T}$ represents the process of removing artificial noise~(e.g., replacing $p_1$ and $v_1$ to $|unk|$ in Figure~\ref{overview_diffusion}), while $X_{t}$ to $X_{t-1}$ represents the process of removing evolutionary noise~(e.g., modifying the $f_{11}$ in the Figure~\ref{overview_diffusion}).
% first challenge
\textbf{To fit the fine-grained editing scenarios}, DivoT5's pre-training takes the old code as input and incorporates token-level \textit{diff} information when generating the complete new code.
Rather than relying on line-level \textit{diff} information to understand code changes, DivoT5 leverages token-level \textit{diff} information to help the model discern which parts of the code need modification and which parts should remain unchanged~(i.e., a mask strategy applied from $X_T$ to $X_{T+1}$ in Figure~\ref{overview_diffusion}).
% Specifically, we design a mask strategy~(applied from $X_T$ to $X_{T+1}$ in Figure~\ref{overview_diffusion}) that focus on tokens remaining unchanged in both input and output, enabling the model to recognize shared content during the editing process.
% second challenge
\textbf{To leverage and enrich the context of the real-world code edits}, we design two additional mask strategies based on real code change histories.
Learning from real code changes means that the model needs to automatically update old versions of code~(which may contain bugs or require new features) to new versions that meet new requirements. 
These two masking strategies~(applied from $X_T$ to $X_{T+1}$ in Figure~\ref{overview_diffusion}) simulate potential issues with code quality or completeness in real code context by introducing varying degrees of random artificial noise into the old code.
This requires the model to understand the context of the specific code and make modifications based on the provided guidance, rather than merely restoring the original version from the corrupted code.
%%%%% Due to the diverse directions of real code changes~\cite{chakraborty2021multi_editing}, natural language~(NL) description is typically used to guide the evolution of the code. By inputting both the NL description and the old code, we can provide rough guidance for the modification of the code with masked context.
% The above three mask strategies ensure that the trained model can understand the required modifications and generate the correct new code across various contexts~(i.e., understanding the evolutionary direction as well as restoration).
% third challenge
\textbf{To learn the characteristics of incremental code evolution}, we design a novel evolutionary direction reinforcement strategy by utilizing intermediate data in the code editing process. 
Intermediate versions~(i.e., $X_t$ ($0 < t \leq T$) in Figure~\ref{overview_diffusion}) of the code that is not fully modified are seen as still containing evolutionary noise.
We utilize intermediate data from the incremental evolution process to train the DivoT5 model, providing clear directional and more detailed guidance information to reinforce the evolutionary direction.
Due to the limitations of applying existing diffusion models to discrete textual data, which typically use multi-step denoising at continued embedding levels and whose performance is constrained by the mapping between continuous embeddings and discrete texts, we apply the above strategies at the data level \zyf{(i.e., conducting the diffusion process on data rather than embeddings)}. 
This model ensures the fluency and reliability of the generated results by directing each version towards the final version~($X_0$), even with the addition of artificial noise~(i.e., constructing training samples $<X_t$, $X_0>$ and generating $X_0$ from any given $X_t$, where $0 < t \leq T+1$). 
% In other words, we follow the same training approach as CodeT5, incorporating the diffusion concept to help the model perceive the real evolutionary direction of code editing, while maintaining the same computational cost during training and inference.
\lqy{
In other words, we adopt the encoder-decoder training framework~\cite{codet52021} while incorporating the diffusion concept to emphasize the evolutionary direction of code editing. Rather than based on the full diffusion model structure, we draw on its style to enhance step-by-step refinement, ensuring the same computational efficiency as the encoder-decoder model~\cite{codet52021,coditt52022,codereviewer} during training and inference.
}

To evaluate DivoT5, we fine-tune DivoT5 on three scenarios~(i.e., NL-guided code editing scenario, code-only editing scenario, and non-editing scenario) using eight benchmark datasets. The evaluation results demonstrate that DivoT5 outperforms the existing state-of-the-art~(SOTA) code pre-trained models for all eight datasets. For example, DivoT5-base~(220M) achieves a significant improvement in exact match performance compared to the CodeT5-base~(220M) model, exceeding a 28.87\% increase on the CodeReview dataset~(from 34.46\% to 44.41\%). 
% Furthermore, DivoT5-base~(220M) outperforms the best-performing code editing pre-trained model~(i.e., CoditT5), increasing performance from 37.19\% to over 44.41\%. 
% Furthermore, DivoT5-base~(220M) significantly surpasses the fine-tuned performance of the 770M model, as well as the few-shot inference results of billion-scale models like DeepSeekCoder-6.7B-instruct, Meta-Llama-3.1-8B-instruct, and ChatGPT.
Furthermore, DivoT5-base~(220M) significantly outperforms the fine-tuned CodeT5-large~(770M)  model and billion-scale models like DeepSeekCoder-6.7B-instruct, Meta-Llama-3.1-8B-instruct, and ChatGPT-gpt-3.5-turbo in few-shot inference.
The small version of DivoT5~(60M) also achieves a good performance in this task, surpassing CodeT5-base~(220M) by 5 percentage points~(from 34.46\% to 40.16\%) in the exact match metric.
In addition, we evaluate the effectiveness of each pre-training task. The ablation results demonstrate the contributions of the designed pre-training tasks to the final performance.

% 我们的贡献
We summarize our contributions as follows:
%%% 数据集
%%% 预训练模型
%%% 实验评估和效果
% ~(\url{https://anonymous.4open.science/r/DivoT5-FSE})
\begin{itemize}
% \vspace{-0.2cm}
    \item  We propose the directional diffusion-style auto-regressive neural model, DivoT5, designed to simulate the characteristics of incremental code evolution carried out by human developers. \textbf{To the best of our knowledge, we are the first to introduce the concept of directional diffusion into the domain of code evolution}.
    % \item  We propose novel pre-training tasks to enrich the context of edited code and learn the incremental editing process. This enables our model to identify fine-grained edits and recognize the direction of code evolution, thereby enhancing its ability to generate evolved code in diverse contexts.
    \item We propose novel pre-training tasks that mimic human-like editing, enabling the model to capture fine-grained edits, track code evolution, and generate updated code across diverse contexts.
    \item  We conduct extensive experiments to evaluate the DivoT5, and the results demonstrate its state-of-the-art performance on various benchmarks related to code evolution.
\end{itemize}

\section{Background and Related Work}
% 代码演化任务
% \vspace{-0.1cm}
\subsection{Code-Editing Tasks and Code-Editing Related Tasks}
During the daily development process, code editing is necessary to maintain the overall functionalities of a software project. In this paper, we differentiate two categories of tasks that are associated with code editing. The first category is tasks involving an editing process to change a code snippet, and we refer to this category as code-editing tasks~\cite{tufano2019coderefine,tufano2021codereview,li2023codeeditor,coditt52022,twinxsql}. The second category is tasks that do not directly edit code but may use code-editing information for other purpose, and we refer to this category as code-editing related tasks~\cite{lin2023cct5,codereviewer}. 
Obviously, it is helpful to automate both categories of tasks during the code maintenance process.

Based on whether natural language guidance is required during the code-editing process, we can further divide code-editing tasks into two types. 
We denote the first type as the \textbf{NL-guided code editing scenario}, such as code refinement~\cite{tufano2021codereview,coditt52022,codereviewer}. Developers provide suggestions for improvement by inspecting and evaluating the code, and the natural language represents the direction for code enhancement. The new code should align with this direction.
We denote the second type as the \textbf{code-only editing scenario}, such as code refinement~\cite{codexglue2021}. For the code refinement task, the new code should address the defects present in the old code. 
In this paper, we primarily consider these two types of code-editing tasks.

% \vspace{-0.2cm}
\subsection{Code Pre-trained Models}
When applying existing pre-training techniques~\cite{bert2018,gpt2018,lewis2019bart,t52020,scis_deeplearning,grammarcoder} to code data, there are two common pre-training tasks, which are referred to as masked language modeling~(MLM)~\cite{codebert2020,codet52021} and denoising auto-encoding~(DAE)~\cite{plbart2021}. 
The basic MLM technique refers to randomly masking a certain proportion of tokens in the input and then predicting the masked tokens.
CodeBERT~\cite{codebert2020} and GraphCodeBERT~\cite{graphcodebert2020} apply the MLM technique to its code data to enhance the model's understanding of code.
% GraphCodeBERT~\cite{graphcodebert2020} also applies the MLM pre-training objective, but it goes beyond by incorporating the learning of code structure information. This includes predicting edges in the data flow graph and establishing alignments between nodes in the source code and the data flow graph.
CodeT5~\cite{codet52021} employs the MLM technique to sequence-to-sequence generation tasks~(referred to as SMLM) to enhance the model's ability to generate code. Additionally, to recognize identifiers within the code, CodeT5 also introduces a variant of SMLM called IMLM~(Identifier Masked Language Modeling). In IMLM, only the identifiers in the source code are masked, while the rest of the tokens are left unchanged, allowing the model to generate code while specifically paying attention to the discernment of identifiers.
DAE is a self-supervised learning approach that enhances model generalization by introducing noise to the input and training the model to reconstruct the original input. In PLBART~\cite{plbart2021}, DAE is introduced during the pre-training stage to train the model in code understanding and generation capabilities. Specifically, PLBART incorporates three noise functions: token masking, token deletion, and token infilling. These functions are applied with a certain probability to disrupt the old code. 
% The model learns the semantics of the input by reconstructing the original input from the perturbed input.

% \vspace{-0.2cm}
% 代码编辑预训练模型
\subsection{Code Editing Pre-trained Models}
%%% 输入代码diff，输出相关预测 code change representation
%%%% 输入旧代码代码及相关信息，输出新代码  code evolution
\qy{Code editing pre-trained models are typically initialized from existing code pre-trained models but are further adapted to specialize in code editing tasks. These models aim to better capture the structure and semantics of code transformations by introducing edit-aware pre-training objectives. Based on code pre-trained models, code editing pre-trained models use the following two ways to consider code edits: \textbf{MLM with code changes} and \textbf{DAE with controlled noises}. }

CodeReviewer~\cite{codereviewer}, CCT5~\cite{lin2023cct5}, and Coeditor~\cite{coeditor_iclr2024} use \textbf{MLM with code changes}. The basic idea of these models is to use a sequence to represent a real-world code change. That is, the output of the \textit{diff} tool when comparing the two code snippets is treated as the sequence for the code change. If there is a comment associated with the code change, the comment is also appended to the sequence. Note that the output of the \textit{diff} represents the line-level differences on two code snippets. On top of the sequences representing code changes, CodeReviewer~\cite{codereviewer} and CCT5~\cite{lin2023cct5} use the standard MLM for pre-training, where the masked parts can be tokens in the following types: tokens to be inserted, tokens to be deleted, tags generated by \textit{diff}, and tokens in the comment. Different from CodeReviewer, CCT5 also adopts GPT-like pre-training tasks and adds data flow information into the sequence representing a code change. 
Similarly, Coeditor~\cite{coeditor_iclr2024} first uses placeholders to mark the sections that need modification, then predicts the next \textit{diff} information based on multiple prior line-level code changes.
% Furthermore, CodeReviewer and CCT5 are trained on different datasets. CodeReviewer is trained on real-world code change data collected by the authors of CodeReviewer. CCT5 is trained on CodeChangeNet, a dataset collected by the authors of CCT5, containing code change data with corresponding code comments.
Our DivoT5 differs from CodeReviewer, CCT5, and Coeditor in two folds. First, DivoT5 uses token-level difference information for pre-training, while CodeReviewer and CCT5 are based on line-level differences. Second, the pre-training tasks in DivoT5 aim to predict code edits on top of the old code (i.e., ensuring evolutionary direction of code editing), thus well aligned with downstream code-editing tasks. However, the pre-training tasks in CodeReviewer, CCT5, and Coeditor do not correspond to any code-editing tasks. For example, the prediction of masked tokens when pre-training CodeReviewer and CCT5 may use information in both code snippets before and after change.

CoditT5~\cite{coditt52022} and CodeEditor~\cite{li2023codeeditor} use \textbf{DAE with controlled noises}. The basic idea of these models is to control noise generation in order that the generated noises can represent the probabilities real-world changes.
% In particular, \textit{Zhang et al}.~\cite{coditt52022} utilize the CodeSearchNet~\cite{csn2019} dataset to design a denoising task for pre-training CoditT5.
In particular, CoditT5~\cite{coditt52022} utilize the CodeSearchNet~\cite{csn2019} dataset to design a denoising task for pre-training. 
Specifically, they calculate the probabilities of inserting, deleting, and modifying tokens in real-world code changes, and use these probabilities to artificially perturb the code. 
Then the standard DAE tasks are adopted for pre-training on top of the perturbed code.
Similar to CoditT5, CodeEditor~\cite{li2023codeeditor} is also based on code perturbed with probabilities from real world. However, the perturbed code is not directly used in pre-training. CodeEditor utilizes an existing code-generation model~\cite{codexglue2021} to further process the perturbed code, and the processed code is then used for pre-training. Thus, the pre-training may well focus on the inserted noises that can hardly be handled by existing models.
Our DivoT5 differs from these models in that DivoT5 is based on predicting real-world code edits while CoditT5 and CodeEditor are based on removing artificially inserted noises. Note that, although CoditT5 and CodeEditor use real-world probabilities to control the insertion of noises, the perturbed code may still largely deviate from real-world human-written code. Furthermore, DivoT5's pre-training tasks also incorporate noise to perturb the old code, but the aim is to help DivoT5 learn to predict code edits in the presence of noise.

\begin{figure*}[t]
  \centering
  % \vspace{-0.13cm}
  % \setlength{\abovecaptionskip}{10pt}
  \includegraphics[scale=1]{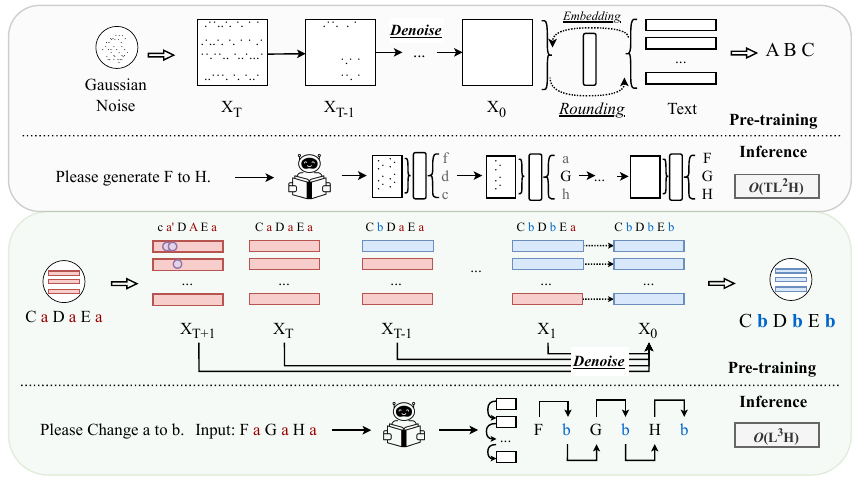}
  \caption{Illustration of traditional diffusion and directional diffusion in text generation. The upper section showcases traditional diffusion, where `ABC' is generated during pre-training, and `FGH' during inference. The lower section highlights directional diffusion, showing `CbDbEb' generated from `CaDaEa' during pre-training, and `FbGbHb' from `FaGaHa' during inference.}
  \label{directionaldiffusion}
  % \vspace{-0.5cm}
\end{figure*}

% \vspace{-0.2cm}
\subsection{Diffusion Models for Text Generation}
% Diffusion models have achieved significant advancements in continuous domains, such as image generation~\cite{diffapp_ho2020denoising}. 
\lqy{Diffusion models, which predominantly operate at the continuous embedding level, have made remarkable progress in continuous domains such as image generation~\cite{diffapp_ho2020denoising}.}
These models involve two key steps: forward noising process and reverse denoising process~\cite{diff_sohl2015deep}.
% diffapp_mittal2021symbolic, diffapp_nichol2021improved, diffapp_kong2020diffwave
Specifically, the forward noise addition process gradually transforms the original data into pure Gaussian noise using a Markov chain, with each noise addition step represented as:
% \vspace{-0.1cm}
\[ q(\mathbf{X_t} | \mathbf{X_{t-1}}) = \mathcal{N} \left( \mathbf{X_t}; \sqrt{1-\alpha_t} \mathbf{X_{t-1}},  \alpha_t \mathbf{I} \right) \]
% \vspace{-0.1cm}
, 
where hyperparameter $\alpha_t$ represents the noise added at diffusion step $t$.
The reverse denoising process gradually removes the noise to generate the target data, with each step represented as:
% \vspace{-0.1cm}
\[ p_{\theta}(\mathbf{X_{t-1}} | \mathbf{X_t}) = \mathcal{N} \left( \mathbf{X_{t-1}}; \mu_{\theta}(\mathbf{X_t}, t), \Sigma_{\theta}(\mathbf{X_t}, t) \right) \]
% \vspace{-0.1cm}
, 
where $\mu_{\theta}$ and $\Sigma_{\theta}$ are learned from the model.

% When applying diffusion models to discrete text data, an additional step is required in the noise addition process to convert discrete text \textbf{W} into continuous embeddings 
\lqy{Existing works applying diffusion models to discrete text data introduce an additional step in the noise addition process, requiring the conversion of discrete text \textbf{W} into continuous embeddings}~(i.e., \( q_\phi (\mathbf{X_0} | \mathbf{W}) = \mathcal{N} (EMB(\mathbf{W}), \alpha_0 \mathbf{I}) \)). Correspondingly, the denoising process includes a trainable rounding step to convert embeddings back into discrete text (i.e., \( p_{\theta}(\mathbf{W} | \mathbf{X_0}) = \prod_{i=1}^N p_{\theta}(W_i | X_i) \))~\cite{difflm_DiffLM_li2022diffusion,difflm_ARDiff_wu2024ar,difflm_Genie_lin2023text}. 
\lqy{
For example, in the upper part of Figure~\ref{directionaldiffusion}, traditional diffusion performs a left-to-right denoising process. The model progressively denoises a continuous embedding and ultimately converts the continuous embedding into discrete token representations through a rounding layer, generating the final text~(e.g., `ABC'). 
Similarly, during the inference process, traditional diffusion also progressively denoises a continuous embedding. Notably, this embedding represents the entire text, and each denoising step generates a new embedding, which is then converted into all tokens of the text~(i.e., in a non-autoregressive manner). As a result, every token in the text has the potential to change at each step of the inference process~(e.g., from pure noise to `fdc', from `fdc' to `aGh', etc.). This implies that achieving coherence and logical consistency across all tokens often requires many iterative steps.
} 

% However, in real-world code evolution scenarios, code typically follows a specific evolution trajectory, and the real noise in code is not in the form of pure Gaussian noise. Moreover, the quality of text generation depends on the embedding and rounding processes, which limits the model's flexibility. The non-auto-regressive~\cite{gu2017nonauto} approach used in the rounding step does not consider the contextual information of the generation process, often resulting in lower generation quality compared to auto-regressive approaches~\cite{transformer2017}. Therefore, this paper proposes a directional diffusion strategy that applies the stepwise denoising on the data level based on real evolution trajectories. Additionally, it incorporates auto-regressive generation to enhance the quality of the generated code.

\lqy{
Our approach diverges from existing methods that apply diffusion to discrete text in two key ways. First, rather than adding noise at the continued embedding level and iteratively removing it, we simulate real-world code editing by leveraging actual code evolution paths. For example, in Figure~\ref{directionaldiffusion}, we treat the partially modified text during the evolution process as ``noise''~(e.g., CbDaEa), while the target of the evolution is considered the ``noise-free'' text~(e.g., CbDbEb). This approach assigns a concrete meaning to each instance of ``noise'', effectively simulating real-world editing scenarios.
Second, we implement a diffusion-style process by utilizing the intermediate state data from the editing process to generate the final noise-free text, where each ``denoising'' step is treated as a transition from a noisy state to a noise-free state. 
% This ensures that the training and inference costs of directional diffusion are consistent with those of traditional text generation models. 
This ensures that we can generate directly usable code in a single step—without multiple denoising iterations—at a computational cost comparable to standard basic models~(e.g., in an autoregressive manner like Transformer~\cite{transformer2017}, CodeT5). 
Specifically, the inference time complexity of traditional diffusion applied to text is about \(O(TL^2H)\), while combining the diffusion concept with traditional text generation models results in a complexity of \(O(L^3H)\), where \(T\) is the number of diffusion steps, \(L\) is the sequence length, and \(H\) is the hidden layer dimension. Typically, \(T\) is significantly larger than \(L\)~(e.g., a sequence of 200 tokens may require up to 2000 denoising steps). Thus, traditional text generation requires less inference time~\cite{difflm_ARDiff_wu2024ar}~(The detailed information in Section\ref{discussion}). 
% This not only ensures more fluent and coherent code outputs but also avoids the embedding and rounding constraints often found in traditional diffusion models, thereby improving both flexibility and overall generation performance.
}

% This means we can generate directly usable code in a single step—without multiple denoising iterations—at a computational cost comparable to standard basic models (e.g., Transformer, CodeT5). 
% Second, we implement our diffusion-style process by manipulating data directly and adopt an autoregressive approach~\cite{transformer2017} for code generation, which contrasts with the non-autoregressive methods~\cite{gu2017nonauto} commonly used in existing diffusion models. 

\begin{figure*}[t]
  \centering
  % \vspace{-0.13cm}
  % \setlength{\abovecaptionskip}{10pt}  % 调整 caption 上方的距离
  \includegraphics[scale=1.2]{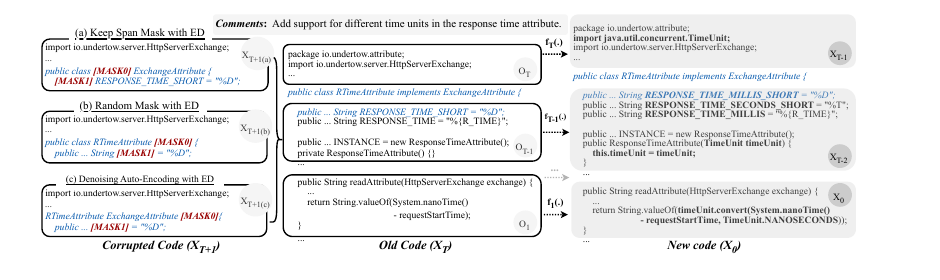}
  % \vspace{-0.3cm}
    \caption{The process of modifying the old code~(center) to fulfill the requirements outlined in the comments~(top left) often necessitates changes across multiple dispersed locations. Achieving the desired evolution (right) requires incremental editing at all relevant points. The blue italics aim to highlight the parts that persist throughout the entire code editing process.}

  \label{overview_example}
  % \vspace{-0.5cm}
\end{figure*}

% \vspace{-0.2cm}
\section{DivoT5}
%In this section, we first present the model architecture and input representation. Then we provide the pre-training data contained the basic editing behavior and explain our edit-aware pre-training objectives. Finally, we present the details of the model implementation details.

DivoT5 is a directional diffusion-style pre-trained model designed to simulate real-world code evolution.
The model leverages the diffusion concept at the data level to capture the incremental nature of code editing and completes specific editing tasks.
% in an auto-regressive manner.
% The model employs a diffusion strategy to incorporates the evolutionary direction and the characteristics of incremental modifications in real programs, and aims to automatically complete specific editing-related tasks using existing natural language or code.

% \vspace{-0.2cm}
\subsection{Model Architecture}
We employ an encoder-decoder framework with an auto-regressive manner based on the Transformer~\cite{transformer2017} model for pre-training. 
Following the existing code editing pre-trained models~\cite{lin2023cct5,coditt52022,codereviewer,liang2024condor}, we continue pre-training DivoT5 on top of the weights of CodeT5~\cite{codet52021}.  
% It allows the model to benefit from the knowledge learned by CodeT5, which can be particularly advantageous in code-related tasks.
This enables the model to leverage CodeT5's knowledge, offering significant advantages in code-related tasks~\cite{coditt52022,chen2024improving,zhao2024spotting}.

% \vspace{-0.2cm}
\subsection{Input Representation}
To accommodate different types of code evolution tasks, our DivoT5 accepts two types of input: input containing only code and input containing both natural language and code. We describe the input of DivoT5 in the following form:
% \vspace{-0.1cm}
$$x  = \{[CLS], w_1, w_2,\cdots,w_K, [SEP], x_1, x_2, \cdots, x_M,[SEP]\}$$
% \vspace{-0.1cm}
, where $w_1, w_2,\cdots,w_K$ represent tokens in natural language, and $x_1, x_2, \cdots, x_M$ represent tokens in the input old code. $[CLS]$ denotes a special token indicating the start of the input, and $[SEP]$ denotes a special separator between different types of inputs.
Similarly, for data that consists solely of code, the input can be simplified to the following form:
$x  = \{[CLS], x_1, x_2, \cdots, x_M,[SEP]\}$.

% \vspace{-0.2cm}
\subsection{Directional Diffusion}
We are the first to introduce the concept of directional diffusion into the domain of code evolution. To better align with the incremental modification characteristic of code evolution, we design a directional diffusion strategy to provide explicit evolutionary direction for the code editing process. 
Figure~\ref{overview_example} shows a specific code example that applies the diffusion concept.

% 1. 代码演化的特性
% 2. 定向diffusion的具体操作
% \vspace{-0.15cm}
\subsubsection{Incremental Modification in Code Evolution}
% why diffusion
% As shown in the code evolution example in Figure~\ref{overview_example}, 
% modifying the old code to meet the requirement described in the comment may involve changes in multiple locations, which could be far apart. Completing this evolution requires incremental modifications to all relevant locations.
As shown in Figure~\ref{overview_example}, modifying the old code~(in the middle) to meet the requirements described in the comments~(at the top left) may involve changes across multiple, potentially distant locations. Successfully completing this evolution~(at the right) requires incremental editing at all relevant points.
In this example, to add support for different time units, the first step is to import the necessary package~(i.e., import the \textit{TimeUnit} package). The next is to redefine variables and implement the corresponding function bodies at the appropriate locations~(i.e., modifying variable \textit{RES\_T\_SHOT} and implementing function \textit{RTime}). The final step is to review and modify all code sections related to the new requirement~(e.g., change the return value of other functions).
Therefore, the code editing process is not a one-off task but requires gradually modifying all relevant code snippets to meet new requirements.
This incremental modification characteristic aligns well with the step-by-step denoising process in diffusion, which prompts us to combine the concept of diffusion with code editing and evolution. 
In the process of code evolution, each modification can be seen as a ``denoising'' step on the old code. 
By applying the diffusion concept to code editing, we are able to simulate a dynamic and step-by-step modification process, allowing the code to incrementally approach the target state with a series of iterations. This strategy not only captures the subtle changes in code evolution but also provides a systematic way to track and predict the improvement path of the code.

% \vspace{-0.15cm}
\subsubsection{Diffusion for the Evolutionary Direction}
% 1. 加噪过程：与已有diffusion的不同；artificial noise + evolutional noise
%%%% - 真实噪音
% 2. 去燥过程：data level not embedding level，auto-regressive

\lqy{Traditional diffusion models handle continuous data (e.g., images) by applying noise to continuous embeddings.}
Existing diffusion models for discrete text generation typically learn text dependencies through a multi-step denoising process at a continuous high-dimensional embedding level and then map these embeddings to natural language text via a rounding process~(e.g., the upper part of Figure~\ref{directionaldiffusion}).
However, compared to natural language, code has a stronger syntactic structure and semantic relationships. Traditional diffusion models might be capable of generating coherent and diverse natural language text through denoising, but generating correct code is much more challenging than generating plain text. 
Furthermore, the real-world code context and code editing process in code evolution scenarios are more difficult to learn using traditional diffusion models.

% Considering the above challenges, we apply a directional diffusion strategy to the data level of code evolution. 
\lqy{
Considering the above challenges, DivoT5, designed for discrete data like code, employs discrete-data-level diffusion, where each noise represents a specific code change, reflecting the natural evolution of code.
}
This strategy allows the entire approach to be completed in an auto-regressive manner as well as to leverage the advantages of existing code-related approaches~\cite{codet52021,lin2023cct5,codereviewer,coditt52022,liang2021lyra,wang2024advanced}.
In the noising stage, we first introduce artificial noise into the old code to learn token-level code editing processes and enrich the context of the old code. 
Then, we treat the intermediate evolution process of editing as a special type of evolution noise to enhance the direction of code evolution. 
Noisy data~(artificial or evolutionary) can be paired with the final new code to form input-output pairs, enabling the model to learn the direction of evolution from different contexts.
In the denoising stage, we treat different types of noise as distinct pre-training tasks, and denoising is performed fully in an auto-regressive manner. This enables the model to consider both the contextual dependencies of code and the incremental characteristics of code evolution. Below, we discuss the noising process at the data level and the denoising process in an auto-regressive manner.

% \vspace{-0.2cm}
\paragraph{Noising process at the data level}
% Code evolution and editing typically involve a clear reference target for the modified code and natural language~(e.g., code comments) to guide the evolution direction.
% The starting point of directional diffusion is not pure Gaussian noise but a snippet of old code that needs modification~($X_T$). 
Code evolution typically requires referencing an existing code, making the starting point of directional diffusion not pure Gaussian noise but an old code that needs modification~($X_T$).
On this basis, we add artificial noise to perturb the old data~($X_{T+1}$), simulating the basic editing process and enriching the context of the old code. 
This approach aims to help the model grasp fine-grained code edits and accurately generate the target new code across various contexts.
We use $X_t$ to represent an intermediate state in the evolution process, indicating that $t$ modifications are still needed to complete all editing operations.
For example, in Figure~\ref{overview_example}, $X_{T-1}$ represents the state where the model has modified the block $O_T$ in the old code $X_T$~(e.g., adding the import of a package). Similarly, $X_{T-2}$ represents the state where the model has modified the block $O_{T-1}$ in $X_{T-1}$~(e.g., changing variables and implementing function bodies).
% To obtain the intermediate version $X_t$, we calculate the \textit{diff} between the old code file and the new code file, and then split the \textit{diff} based on the distance between them.
\qy{
To obtain the intermediate version $X_t$, we compute the \textit{diff} between the old and new code files using Python's \texttt{difflib} library, and then split the \textit{diff} into hunks based on the relative positions of the changes within the code.
}
\qy{Specifically, if transforming old code A (`CaDaEa') to new code B (`CbDbEb') involves three changes (each change representing a \textit{diff} hunk that may affect several nearby lines), the intermediate code evolves incrementally: first modifying the first position to produce `CbDaEa', then applying the second change to produce `CbDbEa'. These intermediate versions follow the edit history sequence, progressing from the beginning of the code to its end and mirroring typical code editing practice.}
We treat the code that has undergone partial human modifications and still requires further modifications as a sample containing evolutionary noise.
By gradually adding artificial noise and evolutionary noise~(human edits) during the code evolution process, we can effectively simulate the transition of code from its evolved state~(i.e., $X_0$) to the corrupted state~(i.e., $X_{T+1}$). 
The process of adding noise can be formalized as follows:
% \vspace{-0.05cm}
\[ q(X_t | X_0) = \sum_{i=1}^t f_{i}(X_{i-1}) \]
% \vspace{-0.05cm}
, where $q(X_t | X_0)$ represents the process of gradually adding noise from $X_0$ to $X_t$, and \( f_{i}(X_0) \) is a $i$-th function representing the addition of noise to \( X_{i-1} \) to obtain \( X_i \). 
For example, in Figure~\ref{overview_example}, \( f_1 \) represents the operation of reverting the modification to the \( O_1 \) block in the evolved code, indicating that there is still a part of the code that needs to be modified. 
Each $X_t$ represents the state of the code at a specific stage of evolution, providing explicit direction that helps the model approach the final state at different modification stages~(i.e., constructing the input-output pairs like $<X_{t}$, $X_0>$). 
This strategy improves the model's adaptability to diverse contexts and enhances its ability to capture subtle changes in the code.

% \vspace{-0.2cm}
% \paragraph{Denoising process within the Encoder-Decoder framework.}
% 已有diffusion
%%% 通过多步（非自回归）去噪学习context
%%% ours：通过mask学习context（基本编辑过程+多样context的生成）并强化方向

% \vspace{-0.15cm}
\paragraph{Denoising process in auto-regressive manner}

Existing diffusion models for discrete text generation typically add and remove noise at the continuous embedding level step by step. 
While this can increase the diversity of the generated text, it requires more iterative steps, making the process slower compared to the mainstream auto-regressive text generation approaches, resulting in lower efficiency.
\lqy{
For example, in the upper part of Figure~\ref{directionaldiffusion}, each denoising step in traditional diffusion models modifies the overall embedding, allowing every token to potentially change in the next step, which facilitates generating diverse text. In contrast, in autoregressive generation, once a token is generated, its content and position are fixed, and the generation continues sequentially from that point onward, facilitating semantic coherence in the output~\cite{transformer2017, codet52021}.
}
Therefore, we conduct the denoising process within an auto-regressive manner similar to CodeT5, using any intermediate state \( X_t \) from the diffusion process as input and producing the final target \( X_0 \) as output. This approach maximizes the generative performance benefits of auto-regression.
% We conduct the denoising process within an Encoder-Decoder framework to capture the internal context and editing processes of the code, thereby maximizing the model's performance.
% To combine the strengths of diffusion and auto-regressive text generation, we propose a new strategy: applying directional diffusion at the data level and integrating it with the auto-regressive generation approach of the Encoder-Decoder model to maximize the model's performance. 
% Specifically, we leverage the auto-regressive nature of the Encoder-Decoder framework, considering contextual information at each generation step to ensure the continuity and accuracy of the generated results.
The denoising process can be formalized as:
% \vspace{-0.05cm}
\[ p_{\theta}(X_{t-1} | X_t) = \prod_{i=1}^{N} p_{\theta}(X_{t-1}^i | X_{t-1}^{1:i-1};X_t) \]
% \vspace{-0.05cm}
, where \( p_{\theta}(X_{t-1} | X_t) \) represents the probability distribution of denoising from \( X_t \) to \( X_{t-1} \), and \( p_{\theta}(X_{t-1}^i | X_{t-1}^{1:i-1};X_t) \) represents the probability of generating the $i$-th token in \( X_{t-1} \), given the existing tokens in \( X_t \) and generated tokens in \( X_{t-1} \).
% The core characteristic of directional diffusion is reflected in each generation step of the auto-regressive model, where the generation target at any step is the final evolution goal $X_0$. 
The core characteristic of directional diffusion is reflected in each generation process of the auto-regressive model, where the generation target at any step is the final evolution goal. In other words, we transform intermediate diffusion data into input-output pairs for the auto-regressive model, such as $<X_{T+1}$, $X_0>$, $<X_T$, $X_0>$, $<X_{T-1}$, $X_0>$.
% Under this strategy, $X_t$ not only represents the state of the code at a specific evolution stage but also clearly points to the final new code $X_0$. 
% Therefore, this approach ensures that each step in the generation process is directional and optimized towards the final new code, resulting in more accurate and consistent outputs from the model in complex contexts.
Under this strategy, each sample in the generation process is directional and optimized towards the final new code, thereby enhancing the model's accuracy and consistency in generating outputs. 
% from the model in complex contexts.
Notably, by converting the directional diffusion process into input-output pairs for the pre-training task, our pre-trained model has the same inference cost as auto-regressive models like CodeT5~(requiring a single inference pass), which is lower compared to the higher inference cost of traditional diffusion models~(requiring multiple inference passes).

% \vspace{-0.2cm}
\subsection{Directional Diffusion Pretraining Tasks}
% DivoT5 aims to learn real-world code editing knowledge to correctly generate new evolved code from old code through four proposed pre-training tasks for directional diffusion.
\qy{
DivoT5 is designed to acquire real-world code editing knowledge by learning to transform old code into completely new code through four directional diffusion-based pre-training tasks.
}
To better illustrate the pre-training tasks, let us assume that the model only has the old code $\textbf{x}$ ($x^1, x^2, \cdots, x^M$) as the input and the new code $\textbf{y}$ ($y^1, y^2, \cdots, y^N$) as the output, where $M$ represents the number of input tokens and $N$ represents the number of output tokens.
Figure~\ref{overview_example} shows an illustration of the four tasks, where the corrupted code on the left~(\(X_{T+1}\)) represents examples of the first three tasks, and the intermediate states~\(X_t\) (\(0 < t \leq T\)) from the old code~(\(X_T\)) in the middle to the new code (\(X_0\)) on the right illustrates the input for the last task.
\qy{
For clarity, we refer to the input code as the old code and the output code as the new code, and use blue italics to highlight the parts of the code that persist throughout the entire editing process.
}
The first three tasks are designed to learn the basic editing processes and contextual dependencies of the code during the diffusion process, while the final task aims to reinforce the direction of code evolution.
We place an additional special token at the prefix of the input sequence to distinguish between different tasks during the training process.
Below we discuss the details of these tasks.

\paragraph{Keep Span Mask with Evolutionary Direction}
% \qy{
% This task aims to make the model aware of which code tokens do not need to be modified~(i.e., Keep Span) during the code editing process. 
% The design of this pre-training task takes into account the token-level similarity in code editing. 
% Since the code before and after evolution typically follows similar patterns or structures, some content may remain unchanged after editing.
% For example, as shown in the $X_{T+1(a)}$ of Figure~\ref{overview_example}, $[MASK1]$ represents the keywords that define the variable~(i.e., `public'), which is typically not changed both before and after the edit.
% }
\minorqy{This task aims to make the model aware of which code tokens do not need to be modified~(i.e., Keep Span) during the code editing process.
The design of this pre-training task leverages the observation that real-world code edits often retain a significant portion of the original structure and content.
Since the code before and after evolution typically follows similar patterns or structures, some tokens remain unchanged across versions.
For example, as shown in $X_{T+1(a)}$ of Figure~\ref{overview_example}, $[MASK1]$ represents the keyword that defines a variable~(i.e., `public'), which commonly remains unchanged during edits.
To leverage this property, we introduce a new masking strategy that explicitly masks unchanged tokens from real code edits, grounding the learning objective in the evolutionary direction.
Unlike generic masking strategies over static code, our task is applied to real-world edit pairs, where each instance reflects a concrete transformation from an earlier version to a later one.
By anchoring the keep span mask in this real edit context~(i.e., evolutionary direction), we encourage the model to learn not only what to change but also what to preserve.
This provides a valuable complementary learning signal: recognizing stable code regions is essential for generating accurate, minimal edits and avoiding unnecessary modifications.
}

% Inspired by IMLM~\cite{codet52021}, where only identifiers in the code are masked, 
% Therefore, we design a new technique in our editing data by masking only the tokens that remain unchanged.
% This allows the model to better learn the common parts frequently shared during code editing, thereby enhancing its ability to understand and edit code. 
% Applying this task of masking tokens to a real-world code editing dataset involves not only masking the shared tokens but also generating new code that aligns with the direction of evolution.

%细节——怎么计算-diff细节
Following existing works~\cite{cu2020,coditt52022}, we calculate the differences between the new code and the old code. However, unlike existing techniques~\cite{lin2023cct5,codereviewer}, our difference information is at the token level rather than at the line level.
Then, we apply the SMLM~\cite{smlm2021} task specifically to the tokens marked as ``KEEP'' in the difference information. 
The formal representation of this task is as follows:
$$\mathcal{L}_{KSM_{ED}}^{\theta} = -\sum_{i=1}^N log \mathcal{P}_{\theta}(y_i|(\textbf{x}^{mask_{KSM}}_{keep}; \textbf{x}_{change};\textbf{y}^{1:i-1}))$$
, where $i$ represents the current generation step, $y_i$ represents the $i$-th token in the new code $y$. The $mask_{KSM}$ represents a type of SMLM approach, $x_{keep}$ indicates the unchanged parts of the code before and after editing, and $x_{change}$ indicates the modified parts.
Since the number of unchanged tokens in this task is smaller than the total number of tokens in the old code, we mask 30\% of the unchanged tokens, aiming to achieve a similar total number of masked tokens as in other tasks.

% \vspace{-0.1cm}
\paragraph{Random Mask with Evolutionary Direction}
\qy{
This pre-training task treats the code editing process as a text generation task with evolutionary direction.
Unlike the previous task, which focuses on unchanged regions, it treats all tokens equally by randomly masking token spans in the old code and learning to generate the completely new code through a sequence-to-sequence framework.
}
% Unlike self-supervised tasks in the field of general text generation, the edit target generation task in code editing involves two processes: generating the masked span and generating the modified code.
Based on the characteristics of the dataset and existing works~\cite{codereviewer,coditt52022}, we set the average span count to 2.5 and the overall percentage of masked tokens to be 20\%.
Unlike self-supervised tasks in the field of general text generation, the goal of this pre-training task is not simply to predict masked words, but rather to generate edited code in the presence of masks (i.e., prediction of the evolutionary direction).
Formally, the loss of this task can be described as follows:
$$\mathcal{L}_{RM_{ED}}^{\theta} = -\sum_{i=1}^N log \mathcal{P}_{\theta}(y_i|\textbf{x}^{mask_{RM}};\textbf{y}^{1:i-1})$$
, where $mask_{RM}$ represents the random mask for the old code.

% \vspace{-0.1cm}
\paragraph{Denoising Auto-Encoding with Evolutionary Direction}
Our third pre-training task combines the characteristics of denoising auto-encoding with the real-world code editing scenarios. It aims to train the model to generate the new code accurately while taking into account the noise and variations that occur in old code. Compared to the previous two pre-training tasks, this task demands a stronger ability for the model to understand code editing. However, this task well aligns with real-world development scenarios because we often cannot guarantee the quality of the old code. It reflects the practical nature of code editing, where the quality and effectiveness of the existing code vary.
Specifically, we apply three types of noise to the old code: insertion, deletion, and replacement~\cite{coditt52022}. As shown in the code snippet $X_{T+1(c)}$ of Figure~\ref{overview_example}, in the old code, we add a $[MASK0]$ token, delete the keyword tokens like $public$, and replace specific variable names with $[MASK1]$. These modifications introduce noise and variations to simulate real-world code editing scenarios. During the generation process, the model not only needs to consider the tokens replaced by $[MASK]$, but also needs to take into account the scenario where existing tokens are deleted and the case of useless masks. 
Unlike existing techniques\cite{coditt52022,li2023codeeditor} that utilize DAE with controlled noises to restore the old code, our approach requires the model to generate the actual new code after removing the introduced noise.
The loss function for this pre-training task is as follows:
% \vspace{-0.1cm}
$$\mathcal{L}_{DAE_{ED}}^{\theta} = -\sum_{i=1}^N log \mathcal{P}_{\theta}(y_i|\hat{\textbf{x}};\textbf{y}^{1:i-1})$$
% \vspace{-0.1cm}
, where $\hat{\textbf{x}}$ represent old code with noise.

% \vspace{-0.1cm}
\paragraph{Evolutionary Direction Reinforcement}
This task aims to reinforce the direction of code editing during the real evolution process. 
We use the final version of the code \(X_0\) as the new code to be generated and any intermediate version of the code \(X_t\)~(where \(0 < t \leq T\)) during the evolution process as the old code input to the model. 
% By learning how to generate new code from old code, the model gains an understanding of code evolution and acquires the ability to edit code.

By incrementally adding details from the old code to the edited code, the model learns the nuances of gradual code modifications and gains the ability to achieve the final desired state of the code.
This task can be formally described as follows:
% \vspace{-0.1cm}
$$\mathcal{L}_{EDR}^{\theta} = -\sum_{i=1}^N log \mathcal{P}_{\theta}(y^i|\textbf{x};\textbf{y}^{1:i-1})$$
% \vspace{-0.1cm}
, where $N$ represents the number of tokens contained in the new code $X_0$, $y^i$ represents the generation to the $i$-th token, and $x$ represents an intermediate version of evolution $X_t$.

% \vspace{-0.1cm}
\subsubsection{Final Loss}
We treat each pre-training task equally and add up the loss values of each task to obtain the overall loss. We describe the final loss function as follows:
% \vspace{-0.05cm}
$$min_{\theta}\mathcal{L}_{Loss}^{\theta} = \mathcal{L}_{KSP_{ED}}^{\theta} + \mathcal{L}_{RM_{ED}}^{\theta} + \mathcal{L}_{DAE_{ED}}^{\theta} + \mathcal{L}_{EDR}^{\theta}$$

% \vspace{-0.2cm}
% % \subsection{Pre-training Details}
% We employ the popular PyTorch~\cite{pytorch} framework for model pre-training on the Linux Ubuntu operating system. Our pre-training process relies on 4 NVIDIA GeForce RTX 3090 GPUs, with each GPU having a memory capacity of 24GB. Additionally, our system has a memory capacity of 251GB. In the small version of pre-training on CodeT5 with 60M parameters, we use a learning rate of 3e-4. For the base version of pre-training on CodeT5 with 220M parameters, we use a learning rate of 2e-4. 
% To expedite the model's pre-training process, we employ the Accelerate~\cite{acclerate} and DeepSpeed~\cite{deepspeed} frameworks with mixed-precision BF16 to improve the training speed.

% \vspace{-0.2cm}
\subsection{Cost of Training and Inference}
We design a directional diffusion strategy at the data level to capture the gradual nature of code editing, transforming diffusion concept into pre-training tasks and direction-guided samples. Then, we train an auto-regressive model, similar to CodeT5, using tasks and samples that incorporate evolution direction to enhance editing capabilities. Therefore, both training and inference consumption are comparable to CodeT5, operating in a single-pass auto-regressive mode, without requiring multiple denoising iterations per sample like traditional diffusion models.

% \vspace{-0.1cm}
\section{Experimental Setup}
% In this section, we first introduce the research questions~(RQs). Then, we introduce the downstream tasks used to evaluate DivoT5. In addition, we present the evaluation metrics and the relevant baseline models used for assessment. Finally, we provide the details of the fine-tuning process in our experiments.

In this section, we first introduce the research questions~(RQs). Then, we introduce the downstream tasks, evaluation metrics, baseline models, and fine-tuning details used to evaluate DivoT5. 

% RQ
% Task & metrics
% Baseline
% - CodeT5
% - CodeReviewer
% - CoditT5
% - CCT5
% \vspace{-0.2cm}
\subsection{Research Questions}
% RQ1: DivoT5 compared with others
% RQ2: DivoT5 generalize
% RQ3: DivoT5 ablation
To evaluate the effectiveness of DivoT5, we ask the following questions:
% \vspace{-0.1cm}
\paragraph{RQ1: How well does DivoT5 perform compared with the existing code editing pre-trained models?} To answer this question, we compare DivoT5 with existing code editing pre-trained models and billion-scale models across multiple downstream tasks.

% \vspace{-0.1cm}
\paragraph{RQ2: How dose DivoT5 generalize to non-editing scenarios?}
To answer this question, we conduct experiments on a code translation task, which differs from the code editing scenarios, to evaluate DivoT5's generalization ability.

% \vspace{-0.1cm}
\paragraph{RQ3: How do the proposed pre-training tasks affect the model performance?}
To answer this question, we pre-train various ablated versions of DivoT5 by removing one pre-training task each, in order to evaluate the effect of each pre-training task. 
\qy{
We conduct these ablation studies on the 60M-parameter variant of DivoT5 to enable efficient experimentation under limited computational resources, following a common practice in prior works~\cite{guo2024deepseekcoder,zhu2024grammart5}.
}

% \vspace{-0.2cm}
\subsection{Pre-training datasets}
\begin{table}[t]
\centering
\renewcommand{\arraystretch}{1.2}
\caption{Statistics of pre-training datasets.}
% \vspace{-0.3cm}
\label{datacompare}
\scalebox{1}{
\begin{tabular}{lll}
\hline
\multicolumn{1}{c}{\textbf{Dataset Name}} & \multicolumn{1}{c}{\textbf{\#Project}}  & \multicolumn{1}{c}{\textbf{\#Data}} \\ \hline
CodeChangeNet~\cite{lin2023cct5}    & 35k+        & 1.5M+    \\
CommitPack-Java~\cite{commitpack_muennighoff2023octopack}    & 130k+     & 3.7M+   \\  \hline
Total   & 165k+     & 5.2M+  \\ \hline
\end{tabular}
}
% \vspace{-0.3cm}
\end{table}
We use real code editing data to train DivoT5, using a code change dataset collected from the open-source repository Github, including CodeChangeNet~\cite{lin2023cct5} and CommitPack~\cite{commitpack_muennighoff2023octopack}. CodeChangeNet includes six popular programming languages~(i.e., Go, Java, JavaScript, PHP, Python, and Ruby), over 35,000 projects, and over 1.5 million commit data. CommitPack comprises 4TB of GitHub commit data, covering 350 programming languages. To conserve computational resources, we select the Java programming language from CommitPack for our experiments. Specifically, the Java dataset includes over 165,000 projects and more than 3.7 million data instances. Generally, more data can lead to better model performance.

Additionally, in the directional diffusion strategy, we focus not only on each data sample but also on the evolution path of that sample. Each intermediate evolution step points towards the final evolved version to reinforce the direction of code editing. Consequently, this strategy increases the amount of pre-training data to nearly 9 million data instances.

% \vspace{-0.2cm}
\subsection{Downstream Tasks}

\begin{table*}[t]
\centering
\renewcommand{\arraystretch}{1.2}
\caption{Statistics of the downstream datasets we used.}
% \vspace{-0.3cm}
\label{datastatistic}
% \scalebox{0.8}{
\begin{tabular}{cc|cccc}
\toprule
\textbf{Scenarios}                                                                         & \textbf{Tasks}                                                                               & \textbf{Datasets}                                & \textbf{\# Train} & \textbf{\# Dev} & \textbf{\# Test} \\ \midrule
\multirow{4}{*}{\textbf{\begin{tabular}[c]{@{}c@{}}NL-guided\\ Code Editing\end{tabular}}} & \multirow{2}{*}{\textbf{\begin{tabular}[c]{@{}c@{}}NL-Based\\ Code Refinement\end{tabular}}} & \textbf{CodeReview}                              & 13,753            & 1,719           & 1,718            \\
                                                                                           &                                                                                              & \textbf{NL-CodeRefinement}                       & 150,406           & 13,103          & 13,104           \\ 
                                                                                           \addlinespace[2pt]\cline{2-6}\addlinespace[2pt]
                                                                                           & \multirow{2}{*}{\textbf{Bug Fixing}}                                                         & \textbf{BugFixing-S}                             & 46,628            & 5,828           & 5,831            \\
                                                                                           &                                                                                              & \textbf{BugFixing-M}                             & 52,324            & 6,542           & 6,538            \\ \midrule
\multirow{3}{*}{\textbf{\begin{tabular}[c]{@{}c@{}}Code-Only\\ Editing\end{tabular}}}      & \multirow{3}{*}{\textbf{Code Refinement}}                                                    & \textbf{Refine-S}                                & 46,680            & 5,835           & 5,835            \\
                                                                                           &                                                                                              & \textbf{Refine-M}                                & 52,364            & 6,545           & 6,545            \\
                                                                                           &                                                                                              & \multicolumn{1}{l}{\textbf{CodeReview-CodeOnly}} & 13,753            & 1,719           & 1,718            \\ \midrule
\textbf{Non-Editing}                                                                       & \textbf{Code Translation}                                                                    & \textbf{CodeTrans}                               & 10,300            & 500             & 1,000            \\ \bottomrule
\end{tabular}
% }
% \vspace{-0.5cm}
\end{table*}

Code editing tasks mainly involve the process of editing old code into new code. During the code editing process, there may be additional constraints that restrict the direction of evolution, such as adhering to natural language requirements. Therefore, the effectiveness of a code editing pre-trained model should be evaluated on two scenarios: the NL-guided code editing scenario and the code-only editing scenario. In particular, we focus on two tasks
~(i.e., the NL-based code refinement task~\cite{tufano2021codereview,coditt52022,codereviewer} and the bug fixing task~\cite{tufano2019learning_bugfixing}) for the first scenario 
and the code refinement task~\cite{tufano2019coderefine,codexglue2021} for the second scenario. We further use the code translation task~\cite{chen2018codetrans,codexglue2021}, which is a non-editing task, for evaluation. Since code translation differs much from our pre-training tasks, the code translation task may help evaluate the generalizability of DivoT5. 
Table~\ref{datastatistic} presents the statistical information of the datasets related to these tasks.

% \vspace{-0.1cm}
\subsubsection{NL-guided code editing scenario}
% Code Review
% bug fixing with more info
%In this type of natural-language-guided code evolution tasks, we focus on the 
% \textbf{The NL-based code refinement task} aims to enhance the old code based on the guidance provided in natural language.
% We first use the CodeReview dataset~\cite{tufano2021codereview} with the Java programming language and adopt the same data settings as \textit{Zhang et al.} \cite{coditt52022}.
% Then, we conduct experiments using a large-scale NL-based code refinement dataset~\cite{codereviewer} that includes nearly 180k data samples across nine programming languages.

\qy{
\textbf{The NL-based code refinement task} focuses on improving existing code based on guidance provided in natural language.
We first conduct experiments on the CodeReview dataset~\cite{tufano2021codereview}, which is written in Java, following the same data setup as \textit{Zhang et al.}\cite{coditt52022}.
To further evaluate the model's capability, we also utilize a large-scale multilingual dataset\cite{codereviewer}, which contains nearly 180k samples spanning nine programming languages.
\textbf{The bug fixing task} aims to fix the bug existing in the old code.
We use the bug-fixing dataset with additional information~\cite{tufano2019learning_bugfixing,chakraborty2021multi_editing}, such as developers' natural language guidance, to help the model fix bugs more accurately.
}

% \vspace{-0.1cm}
\subsubsection{Code-only editing scenario}
% code refinement
% translation
%In code-only evolution tasks, we focus on two types of tasks: code refinement and code translation tasks.

\textbf{The code refinement task} aims to convert defective code into correct one without natural language guidance. 
We use the Java dataset collected by \textit{Tufano et al.}~\cite{tufano2019coderefine} to evaluate the model's performance on this task. This dataset comprises two versions of benchmarks: the Refine-small version with code tokens fewer than 50, and the Refine-medium version with code tokens ranging between 50 and 100. We follow the settings of CodeXGLEU~\cite{codexglue2021} and use the same metrics to evaluate our model.
\qy{To more comprehensively evaluate this scenario, we construct a variant of the CodeReview~\cite{tufano2021codereview} dataset, referred to as CodeReview-CodeOnly, by removing the natural language input. This version retains the same number of samples as the original dataset but omits the NL comments from the input, enabling us to isolate and assess the model’s performance in the absence of natural language guidance.}

% \vspace{-0.1cm}
\subsubsection{Non-editing scenario}
\textbf{The code translation task} involves translating a program written in one programming language into another programming language. This task assists developers in code migration and cross-platform development using different programming languages. We use the CodeTrans dataset~\cite{chen2018codetrans,codexglue2021} as our benchmark for this task, which includes Java and CSharp code samples, and evaluate it with the same metrics as the code refinement task.
% We evaluate the model using the same metrics as used in the code refinement task.

% \vspace{-0.2cm}
\subsection{Metrics}
We follow the existing research in the field of code generation~\cite{codexglue2021} and code evolution~\cite{coditt52022} and select four metrics to evaluate the generated code: Exact Match, BLEU, CodeBLEU, and SARI. 
% Below we describe the details.

% \vspace{-0.1cm}
\subsubsection{Exact Match (EM)}
The EM score~\cite{Xinyun2018EM}, often used in natural language processing tasks such as question answering, evaluates the accuracy of model predictions by measuring the proportion of examples for which the predicted answer exactly matches the ground truth answer. 
% The equation for calculating EM is as follows:
% \begin{equation*}
% \text{EM} = \frac{1}{N} \sum_{i=1}^{N} \text{I}(\hat{y}_i = y_i)
% \end{equation*}
% $\hat{y}_i$ is the predicted answer for example $i$. $y_i$ is the ground truth answer for example $i$. $N$ is the total number of examples. $\text{I}(\cdot)$ is the indicator function, which returns 1 if the condition within the brackets is true and 0 otherwise.

% \vspace{-0.1cm}
\subsubsection{Bilingual Evaluation Understudy (BLEU)}
The BLEU~\cite{Kishore2002BLEU} score, also commonly used in natural language processing, evaluates the accuracy of model predictions by measuring the ratio of n-gram matches between the generated answer and the ground truth answer. 
% The equation for calculating BLEU is as follows:
% \begin{equation*}
% \text{BLEU} = \text{BP} \times \exp\left(\sum_{n=1}^N \frac{1}{N} \log p_n\right)
% \end{equation*}
% $\text{BP}$ is the brevity penalty, which penalizes shorter generated answers. $N$ is the maximum n-gram length considered. $p_n$ is calculated as the ratio of n-gram matches in the generated answer to the total number of n-grams in the generated answer. In practice, the BLEU score is often reported as a percentage, with a higher BLEU score indicating better quality of the generated answer.
In practice, the BLEU-4 score is often reported, with a higher BLEU-4 score indicating better quality of the generated answer. 

% \vspace{-0.1cm}
\subsubsection{CodeBLEU (C-BLEU)}
The CodeBLEU~\cite{Shuo2020CodeBLEU} score, adapted from BLEU and designed specifically for programming language processing, evaluates the accuracy of model predictions by summing four weighted parts. 
The equation for calculating CodeBLEU is as follows:
\begin{align*}
\text{C-BLEU} &= \alpha\cdot\text{BLEU} + \beta\cdot\text{BLEU}_{\text{weight}}\nonumber+ \gamma\cdot\text{Match}_{\text{ast}} \\ + \delta\cdot\text{Match}_{\text{df}}
\end{align*}

% $BLEU$ is calculated as described in the last subsection. 
% , where $BLEU$ is described in the last subsection. 
, where $BLEU_{weight}$ is the weighted $BLEU$, obtained by comparing the generated answer and the ground truth answer with different weights on different tokens. $Match_{ast}$ is the syntactic AST match. $Match_{df}$ is the semantic dataflow match. $BLEU_{weight}$ and $Match_{ast}$ are used to measure grammatical correctness, and $Match_{df}$ is used to calculate logic correctness.

\subsubsection{System output Against References and Input (SARI)}
\qy{
Evaluating code editing models requires not only output similarity but also an accurate assessment of how well the model performs actual edit operations~\cite{editprogress}. 
The SARI~\cite{sari} metric, originally proposed for text simplification, evaluates the quality of edits by comparing the system output against both the input and the reference. Specifically, it separately measures the precision and recall of three types of operations: additions, deletions, and keeps. 
Unlike similarity-based metrics such as BLEU or CodeBLEU, SARI explicitly focuses on the correctness of edit operations, making it more suitable for code editing tasks.
}

% \vspace{-0.2cm}
\subsection{Baselines}
Despite the availability of numerous large-scale code pre-trained models~\cite{chatgpt2024,guo2024deepseekcoder}, fine-tuning is challenging due to their massive parameter sizes. 
Therefore, we compare the fine-tuned performance of models with similar and larger parameter sizes and the few-shot performance of billion-scale instruct models.
\lqy{
% Traditional diffusion models are designed for continuous data~(e.g., images) by applying noise to continuous embeddings. However, applying embedding-level noise to code can result in drastic and unrealistic changes, failing to capture the incremental nature of code evolution. Additionally, while diffusion methods have been explored in text generation, producing coherent text using such models is computationally intensive, and generating correct code demands even more processing steps. Due to computational constraints, we aim to fine-tune DivoT5 to use the same resources as other models~(e.g., CodeT5), so no comparisons with traditional diffusion models are included. 
Since our approach draws on the diffusion style to simulate the step-by-step editing process in real-world development scenarios, we focus our comparisons on auto-regressive code models rather than traditional diffusion models. By referencing diffusion concepts to generate data for training, our approach adapts these principles within an auto-regressive framework, making direct comparisons with traditional diffusion models~(i.e., non-autoregressive framework) outside the scope of this work.
}

% Below we introduce the corresponding baselines for each scenario.
% We categorize these models into three groups: code pre-trained models, code editing pre-trained models, and billion-scale models. 
% For the code pre-trained models, we select the widely used CodeT5~\cite{codet52021} and CodeT5+~\cite{wang2023codet5+} as our baselines. As for the code editing pre-trained models, we select CodeReviewer~\cite{codereviewer}, CCT5~\cite{lin2023cct5}, CoditT5~\cite{coditt52022}, and CodeEditor~\cite{li2023codeeditor} as our baselines.

% \subsubsection{Baselines for NL-guided code editing scenario}
% We chose CodeT5, CodeT5+, CodeReviewer, CCT5, CoditT5, and CodeEditor as baselines for each task in this scenario.

% If the original papers provide performance metrics, we use the data from those papers. We fine-tune the models released in the original papers on specific datasets to obtain the corresponding metrics. If the baseline models are not publicly available or if the original papers provide only partial metrics, we indicate this with a '-' in the table.

% \subsubsection{Baselines for Code-only editing scenario}

% \subsubsection{Baselines for Non-editing scenario}
% \vspace{-0.1cm}
\subsubsection{Code pre-trained models}
CodeT5 is a pre-trained model based on the encoder-decoder structure of T5~\cite{t52020}.  
During the pre-training process, CodeT5 applies four pre-training tasks: masked span prediction, identifier tagging, masked identifier prediction, and bimodal dual generation. Identifier tagging is used to label whether a token is a code identifier, while the masked identifier prediction task is used to generate the masked identifier content.
Experimental results demonstrate that CodeT5 can better capture code semantics and outperforms previous approaches on code-related tasks. In our evaluation, we use three variants of CodeT5: CodeT5-small (60M), CodeT5-base (220M), and CodeT5-large (770M).
Compared to CodeT5, CodeT5+ introduces more pre-training data and tasks, such as span denoising, contrastive learning, and causal LM. In our evaluation, we use the 220M parameter version of CodeT5+ (which is of the same scale as DivoT5-base) for comparison.
% \qy{
% CodeLlama~\cite{codellama} is a series of decoder-only models pre-trained on a large corpus of natural and programming languages, with versions specifically optimized for code understanding and generation. It has demonstrated strong performance across various code-related tasks. The 6.7B parameter version of CodeLlama~(base) is used in our experiments, which serves as a powerful baseline for comparison against DivoT5.
% }
\minorqy{CodeLlama~\cite{codellama} is a family of decoder-only models pre-trained on large-scale corpora of both natural language and code, with variants specifically optimized for code understanding and generation. It has shown strong performance across a wide range of code-related benchmarks. In our experiments, we use the 6.7B base version of CodeLlama as a representative and competitive LLM baseline to compare against DivoT5, highlighting the advantages of our task-specific pretraining approach.}

% \vspace{-0.1cm}
\subsubsection{Code editing pre-trained models}
In our evaluation, we use four code editing pre-trained models: CodeReviewer, CCT5, CoditT5, and CodeEditor. Among the four models, CodeEdit has 60M parameters, and the other three have 220M parameters.
CodeReviewer is a pre-trained model on a real-world dataset in code change scenarios. The primary pre-training task of CodeReviewer is to utilize a masked language modeling approach to mask the sequence representing line level code differences, and then predict the masked content.
Additionally, CodeReviewer also incorporates a pre-training task of generating commit messages from code differences to enhance the model's ability to generate text.
CCT5 is a pre-trained model on a large dataset consisting of 1.5M+ pairwise data of code changes and commit messages from popular GitHub projects. 
In the pre-training process, CCT5 leverages the difference sequence of code and applies the SMLM task to learn the representation of code changes. For example, CCT5 first masks the information in the difference sequence at the granularity of code lines, then use other information from the difference and the commit message to predict the masked content. 
CoditT5 designs denoising pre-training tasks specifically for code editing. CoditT5 first calculates the probability of code edits in downstream tasks, including the probabilities of token insertion, deletion, and modification. Based on these probabilities, CoditT5 introduces noise to the old code. Finally, the model learns the code editing process by reconstructing the old code from the perturbed code. 
CodeEditor first disrupts the old code based on probabilities related to editing operations. Then, CodeEditor utilizes the existing generation model~\cite{codexglue2021} to fill in the disrupted parts as input. Similar to CoditT5, CodeEditor also employs a denoising approach to recover the old code from the disrupted code. 

% \vspace{-0.1cm}
\subsubsection{Instruct-Models with billion-scale parameters}
% \qy{
% To more comprehensively evaluate the contribution of our model, we compare its performance against several billion-scale instruct models that have been fine-tuned for dialogue settings, including ChatGPT~(gpt-3.5-turbo)\cite{chatgpt2024}, DeepSeekCoder-6.7B-instruct\cite{guo2024deepseekcoder}, and Meta-Llama-3.1~(8B-instruct)~\cite{llama3}. These models are widely adopted for their strong generative capabilities in general-purpose code generation tasks. Due to the high computational cost of fine-tuning such large models, we evaluate them using a few-shot setting instead. For the few-shot prompts, we refer to existing work~\cite{chatgpt_refine}, providing scenario descriptions of the problems and setting a low temperature to obtain the most stable and accurate results possible. 
% }
\minorqy{
To more comprehensively evaluate the effectiveness of our model, we compare it against several billion-scale instruction-tuned LLMs that are widely used in code generation and editing tasks~\cite{chatgpt2024,guo2024deepseekcoder,llama3}. These include ChatGPT~(gpt-3.5-turbo)~\cite{chatgpt2024}, DeepSeekCoder-6.7B-Instruct~\cite{guo2024deepseekcoder}, and Meta-Llama-3.1~(8B-Instruct)~\cite{llama3}. These models are known for their strong general-purpose code generation capabilities and are representative of current state-of-the-art instruction-tuned LLMs.
Given the high computational cost of fine-tuning such large models on our task, we evaluate them in a few-shot setting instead. Following prior work~\cite{chatgpt_refine}, we design prompts that include scenario-specific descriptions of the editing task and use a low temperature to ensure consistent and accurate outputs.
}
\lqy{
Specifically, we refine prompts several times based on ~\cite{chatgpt_refine}, with a temperature of 0 for reproducibility, and choose a top-performing few-shot example from the training set after 5 attempts.
All the prompts used in our experiments can be found in~\cite{divot5}.
}
Additionally, since these models are not fine-tuned on the training set, to ensure a fairer comparison, we remove the code style elements such as casing, line breaks, and comments from both the model-generated results and the ground truth in the test set.
However, tasks related to code editing have strong domain-specific characteristics, which often limit the performance of these models.

% \vspace{-0.3cm}
\subsection{Training Details}
In the pre-training phase, we employ the popular PyTorch~\cite{pytorch} framework for model pre-training on the Linux Ubuntu operating system. Our pre-training process relies on 4 NVIDIA GeForce RTX 3090 GPUs, with each GPU having a memory capacity of 24GB. Additionally, our system has a memory capacity of 251GB. In the small version of pre-training on CodeT5 with 60M parameters, we use a learning rate of 3e-4. For the base version of pre-training on CodeT5 with 220M parameters, we use a learning rate of 2e-4. 
To expedite the model's pre-training process, we employ the Accelerate~\cite{acclerate} and DeepSpeed~\cite{deepspeed} frameworks with mixed-precision BF16 to improve the training speed.

During the fine-tuning process for downstream tasks, we maintain the same system environment and acceleration settings as in the pre-training phase.
We set length restrictions for the inputs and outputs of each task based on a statistical analysis of the dataset samples and existing works~\cite{coditt52022,codexglue2021,codereviewer}. All models use the same length settings for the same dataset.

% \vspace{-0.2cm}
\section{Results}
% In this section, we provide answers to each research question and present our experimental results in detail.

% Answer to research questions
% Effectiveness of DivoT5
% \vspace{-0.1cm}
\subsection{RQ1: Effectiveness of DivoT5}
\qy{
We evaluate the performance of DivoT5 on two types of code editing scenarios~(i.e., NL-guided code editing scenario and code-only editing scenario) across seven specific datasets, comparing it with existing code pre-trained models.
If the model has already been evaluated on the target datasets, we directly report the performance metrics from the original paper. Otherwise, we download the released model from HuggingFace~\cite{huggingface} and use the official code to evaluate it on the corresponding datasets. For models whose checkpoints or code are not publicly available, we only include a comparison with our results on the same dataset, and mark missing entries with a `-' in the table. 
% We use $^\alpha$ to indicate cases where a model shows no statistically significant difference from DivoT5 under a specific metric.
We use the symbol $^\alpha$ in the result tables to indicate models whose performance is not statistically significantly different from DivoT5, with both DivoT5 and the corresponding baselines jointly marked to reflect their comparable results.
}

% \vspace{-0.1cm}
\subsubsection{NL-guided code editing scenario}

\begin{table*}[t]
\centering
\renewcommand{\arraystretch}{1.1}
\caption{Results of the NL-guided code editing scenario~(the FS represents few-shot manner).}
% \vspace{-0.3cm}
\label{codereview}
\scalebox{0.97}{
\begin{tabular}{lccc|ccc|ccc|ccc}
\hline
\textbf{Models}                  & \multicolumn{3}{c|}{\textbf{CodeReview}}         & \multicolumn{3}{c|}{\textbf{\begin{tabular}[c]{@{}c@{}}NL-\\ CodeRefinement\end{tabular}}} & \multicolumn{3}{c|}{\textbf{BugFixing-S}}        & \multicolumn{3}{c}{\textbf{BugFixing-M}}         \\ \hline
\textbf{Metrics}                 & \textbf{EM}    & \textbf{BLEU}  & \textbf{SARI}  & \textbf{EM}                  & \textbf{BLEU}                & \textbf{SARI}                & \textbf{EM}    & \textbf{BLEU}  & \textbf{SARI}  & \textbf{EM}    & \textbf{BLEU}  & \textbf{SARI}  \\ \hline
CodeT5-small (60M)               & 32.25          & 85.86          & 69.79          & 28.07                        & 78.06                        & 69.82                        & 33.10          & 73.97          & 58.45          & 25.45          & 85.01$^\alpha$          & 55.62          \\
CodeT5-base (220M)               & 34.46          & 86.21          & 70.94          & 29.44                        & 78.64                        & 70.61                        & 36.84          & 75.45          & 59.85          & 30.47          & 85.68$^\alpha$          & 57.10          \\
CodeT5-large (770M)              & 34.63          & 86.10          & 71.28          & 29.81                        & 79.17$^\alpha$                        & 71.07                        & 36.55          & 75.50$^\alpha$          & 59.56          & 29.83          & \textbf{85.86} & 57.03          \\
CodeT5+ (220M)                   & 32.36          & 85.99          & 69.71          & 30.24                        & 78.64                        & 71.07                        & 34.27          & 73.88          & 58.87          & 25.19          & 84.66          & 55.22          \\
CodeEditor (60M)                 & 29.40          & -              & -              & -                            & -                            & -                            & -              & -              & -              & -              & -              & -              \\
CoditT5 (220M)                   & 37.19          & 86.58          & 70.33          & 28.95                        & 78.98$^\alpha$                        & 70.62                        & 37.52          & 75.39          & 59.72          & 29.96          & 85.75          & 56.35          \\
CCT5 (220M)                      & 36.96          & 87.65          & 70.25          & 29.66                        & 78.55                        & 70.48                        & 36.72          & 74.95          & 59.64          & 30.15          & 84.78          & 57.02          \\
CodeReviewer (220M)              & 37.14          & 87.30          & 71.21          & 30.67                        & 78.96$^\alpha$                        & 71.49                        & 36.36          & 74.28          & 59.57          & 31.16          & 85.34$^\alpha$          & 57.05          \\
CodeLlama (6.7B-base)            & 31.49          & 85.19          & 67.44          & 31.65                        & 79.07$^\alpha$                        & 68.64                        & \textbf{40.44} & 74.36          & \textbf{60.94}$^\alpha$ & 29.40          & 83.91          & 56.62          \\ \hline
DeepSeekCoder-FS (6.7B-instruct) & 26.83          & 76.17          & 66.26          & 15.78                        & 70.04                        & 60.49                        & 2.68           & 43.76          & 47.59          & 2.26           & 59.63          & 46.31          \\
Meta-Llama-3.1-FS (8B-instruct)  & 30.15          & 71.39          & 66.91          & 18.28                        & 65.45                        & 61.59                        & 2.30           & 34.02          & 47.25          & 2.26           & 47.54          & 46.14          \\
ChatGPT-FS (gpt-3.5-turbo)       & 34.81          & 82.05          & 68.39          & 21.99                        & \textbf{80.47}               & 65.24                        & 2.42           & 54.98          & 48.07          & 1.71           & 65.71          & 47.43          \\ \hline
DivoT5-small (60M)               & 40.16          & 87.81          & 72.00          & 29.42                        & 78.59                        & 70.93                        & 35.84          & 74.80          & 59.40          & 28.72          & 84.65          & 56.78          \\
DivoT5-base (220M)               & \textbf{44.41} & \textbf{88.54} & \textbf{73.51} & \textbf{32.75}               & 79.19$^\alpha$                        & \textbf{72.09}               & 39.08          & \textbf{75.86}$^\alpha$ & 60.67$^\alpha$          & \textbf{33.65} & 85.24$^\alpha$          & \textbf{58.12} \\ \hline
\end{tabular}
}
% \vspace{-0.3cm}
\end{table*}

% \subsubsection{Automated Code Review Task}
% We choose the automated code review task to evaluate the performance of DivoT5 on natural-language-guided code evolution tasks.
Table~\ref{codereview} presents a comparison of the performance for the NL-guided code editing scenario, emphasizing the effectiveness of DivoT5. 
It can be observed that the DivoT5-base model demonstrates significant improvements across most datasets compared to all other models, including 60M, 220M, 770M, and billion-scale parameters.

\qy{
For the NL-based code refinement task, DivoT5-base achieves a substantial improvement over CodeT5-base, with a 29\% relative gain in the exact match metric (from 34.46 to 44.41), and consistently stronger results across BLEU and SARI. Notably, the smaller DivoT5-small model even surpasses models with 220M parameters in all evaluation metrics, demonstrating the effectiveness of our pretraining objectives even at a smaller scale. Compared to other pre-trained editing models such as CoditT5 and CodeReviewer, DivoT5-base also achieves over 18\% higher performance in the exact match metric, further validating its superior ability to model real-world code edits.
}
\qy{
In the large-scale, multilingual NL-CodeRefinement dataset, which spans nine programming languages, DivoT5-base continues to outperform strong baselines. It achieves 32.75 in exact match and 72.09 in SARI, both improvements over CodeT5-base, which reaches 29.44 and 70.61 respectively. These results highlight DivoT5’s robustness and generalization capability across diverse codebases and natural language annotations.
Since the fine-tuned weights of CodeT5-base from~\cite{codereviewer} are unavailable, we re-fine-tuned CodeT5 on this dataset for 13 epochs over 44 hours and 52 minutes until performance plateaued on the validation set.
}

\begin{figure}[t]
  \centering
  \includegraphics[scale=0.375]{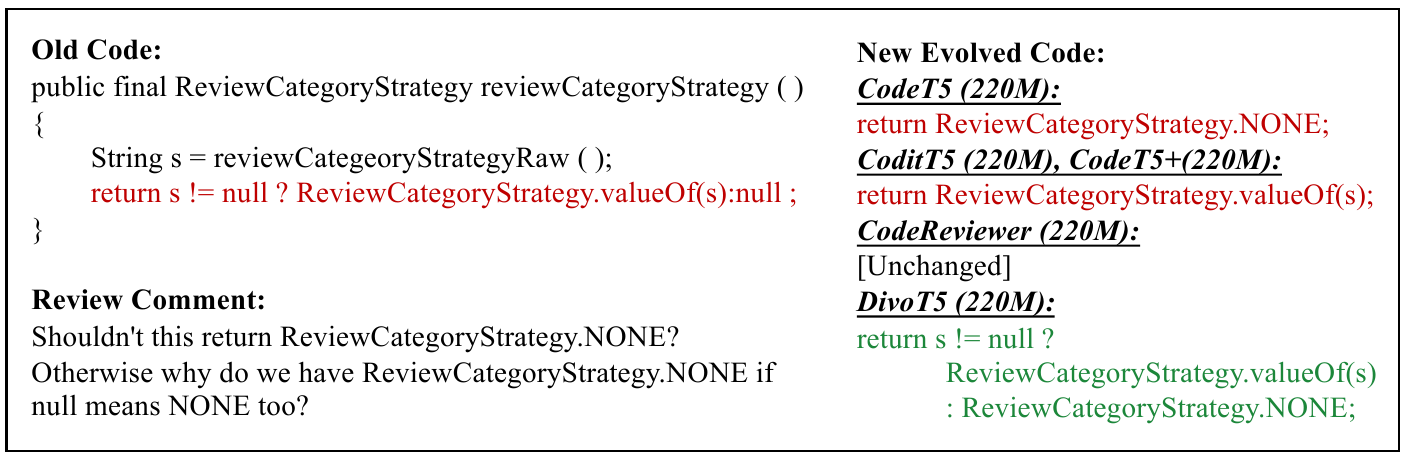}
  % \vspace{-0.3cm}
  \caption{Comparison of generation case using different models on the CodeReview dataset}
  \label{case_cr}
  \vspace{-0.51cm}
\end{figure}

Figure~\ref{case_cr} depicts a case study showcasing different models' outputs on the CodeReview dataset. To save space, we only compare the outputs of models with a parameter count of 220M and display the differences between the new code and the old code. 
In the Figure~\ref{case_cr}, the old code's functionality checks for null in the return value. The review comment suggests using ``ReviewCategoryStrategy.NONE'' instead of returning a null value. 
Specifically, the reviewer recommends replacing the return value in the last line of the old code but not changing the null value used for checks.
The CodeT5 model returns the ``ReviewCategoryStrategy.NONE'' value, incorrectly removes the necessary condition check. 
CoditT5 and CodeT5+ mistakenly delete the original null check, while the CodeReviewer model outputs the old code without making any modifications.
In this example, DivoT5 is able to generate the evolved new code correctly. This could be attributed to the design of DivoT5's pre-training task, which enables the model to perceive which parts do not require modification~(i.e., the $KSM_{ED}$ pre-training task). Notably, despite being trained only on Java data from the CommitPack dataset due to resource constraints, DivoT5 achieves a higher EM score than the CodeReviewer model, which was trained on data from multiple programming languages.

\qy{
Similarly, for the bug fixing task, DivoT5 outperforms all models of comparable size across most evaluation metrics. On the BugFixing-S dataset, CodeLlama (6.7B-base) achieves the highest scores with 40.44 in exact match and 60.94 in SARI, demonstrating strong performance on short, localized edits. However, DivoT5-base closely follows with 39.08 in exact match and 60.67 in SARI, despite having over 30× fewer parameters. This result underscores the effectiveness of our fine-grained edit-oriented pretraining in achieving competitive performance without relying on massive model scale.
On the more challenging BugFixing-M dataset, which involves longer and more complex edits, DivoT5-base further strengthens its lead with 33.65 in exact match and 58.12 in SARI, outperforming all other models, including those with significantly larger parameter counts. 
While DivoT5-base shows a slight decrease in BLEU score compared to CodeT5-base in the BugFixing-M dataset, it increases EM by over 10\%. Given that the EM metric is typically more crucial than the BLEU score in indicating correct bug fixes, this improvement is significant. 
}

When compared to billion-scale models in few-shot settings, we observe that these models generally show limited performance in NL-guided code editing scenarios. This is because code editing tasks are highly domain-specific, and the few-shot approach often struggles to learn sufficient domain knowledge. This observation aligns with existing research findings~\cite{chatgpt_refine}. Additionally, in NL-based code refinement, the natural language descriptions are more closely related to the actual code changes, resulting in better performance from few-shot models. However, for bug-fixing tasks, the natural language descriptions typically provide relevant context rather than directly corresponding to the actual changes, leading to poorer performance from these models.

% \vspace{-0.1cm}
\subsubsection{Code-only editing scenario}
\qy{
Table~\ref{refinement} summarizes the performance of DivoT5 and various baselines in the code-only editing scenario, which is evaluated across three representative datasets: Refine-S, Refine-M, and CodeReview-CodeOnly. In this setting, models are required to refine code without access to natural language guidance. Due to the substantial lexical overlap between old and new code versions, BLEU scores tend to be uniformly high and less discriminative. Therefore, we primarily emphasize the EM and SARI metrics, which more directly reflect the model’s ability to perform accurate and meaningful edits.
}

\qy{
On the Refine-S dataset, DivoT5-small (60M) performs on par with or better than all 220M-parameter baselines, achieving 22.31 in EM and 47.38 in SARI. The DivoT5-base model (220M) achieves the highest overall performance, with 23.02 in EM and 47.92 in SARI, surpassing CodeT5-base and CodeReviewer by over 6\% in exact match. These results demonstrate that DivoT5 learns effective editing strategies even without natural language cues, outperforming larger models with fewer parameters.
}

\qy{
On the more challenging Refine-M dataset, which involves longer and more complex code segments, DivoT5-base continues to lead with 16.55 in EM and 43.09 in SARI, improving upon CodeT5-base by 2.59 percentage points in EM. While BLEU scores remain high (around 88) for all models due to large unchanged regions, DivoT5 still achieves competitive fluency (88.58 in BLEU), showing no trade-off between correctness and surface similarity.
}

\qy{
On the CodeReview-CodeOnly dataset, which is constructed by removing the natural language guidance from the original CodeReview dataset, DivoT5 demonstrates even more pronounced advantages. Specifically, DivoT5-base achieves a 34.3\% relative improvement in EM and a 10.9\% relative improvement in SARI compared to CodeReviewer. In comparison to CodeT5+, DivoT5-base also shows clear gains, with EM increasing from 12.75 to 20.90 and SARI improving from 39.42 to 46.85. These results highlight the effectiveness of DivoT5 in code-only editing scenarios, particularly in the absence of external natural language signals.
}

\qy{
Due to the lack of sufficient domain knowledge and guidance from natural language on the direction of changes, billion-scale few-shot models such as ChatGPT, Meta-Llama-3.1, and DeepSeekCoder perform poorly across all datasets. Although they often achieve reasonable BLEU scores by generating syntactically similar code, they fail to produce exact outputs or apply semantically faithful edits. The knowledge of edit directions and patterns required by these datasets often needs to be acquired through fine-tuning the training set.
}

\lqy{The effectiveness of DivoT5 stems from its ability to bridge the significant information gap when evolving code. For example, when editing from code A (`CaDaEa') to code B (`CbDbEb')), instead of directly generating B from A, we introduce intermediate states, like `CbDaEa' and `CbDbEa', enabling smoother transitions from A to B, enriching the potential evolution paths to alleviate the difficulty of direct generation. 
}

\begin{table*}[t]
\centering
\renewcommand{\arraystretch}{1.1}
\caption{Results of the code-only editing scenario.}
% \vspace{-0.3cm}
\label{refinement}
\scalebox{1}{
\begin{tabular}{lccc|ccc|ccc}
\hline
\textbf{Models}                  & \multicolumn{3}{c|}{\textbf{Refine-S}}           & \multicolumn{3}{c|}{\textbf{Refine-M}}           & \multicolumn{3}{c}{\textbf{CodeReview-CodeOnly}} \\ \hline
\textbf{Metrics}                 & \textbf{EM}    & \textbf{BLEU}  & \textbf{SARI}  & \textbf{EM}    & \textbf{BLEU}  & \textbf{SARI}  & \textbf{EM}    & \textbf{BLEU}  & \textbf{SARI}  \\ \hline
CodeT5-small (60M)               & 19.06          & 76.23          & 44.69          & 10.92          & 89.20          & 38.73          & 10.77          & 82.18$^\alpha$          & 38.27          \\
CodeT5-base (220M)               & 21.61          & 77.43$^\alpha$          & 45.78          & 13.96          & 87.64          & 40.25          & 13.15          & 82.56$^\alpha$          & 39.54          \\
CodeT5-large (770M)              & 21.7           & 77.38$^\alpha$          & 46.03          & 14.76          & 89.22          & 41.08          & 13.34          & 82.58$^\alpha$          & 39.87          \\
CodeT5+ (220M)                   & 22.18          & \textbf{78.27} & 46.87          & 15.13          & 88.64$^\alpha$          & 42.40          & 12.75          & 82.67$^\alpha$          & 39.42          \\
CoditT5 (220M)                   & 18.75          & 77.59$^\alpha$          & 44.61          & 13.00          & \textbf{89.45} & 40.15          & 14.09          & 83.13          & 41.04          \\
CCT5 (220M)                      & 17.87          & 77.02          & 43.25          & 9.75           & 89.06          & 38.21          & 13.56          & 82.31$^\alpha$          & 41.59          \\
CodeReviewer (220M)              & 21.66          & 77.50$^\alpha$          & 45.89          & 12.61          & 88.92$^\alpha$          & 39.57          & 15.56          & 81.97          & 42.26          \\
CodeLlama (6.7B-base)            & 20.17          & 77.24$^\alpha$          & 45.78          & 13.23          & 88.30$^\alpha$          & 40.96          & 14.90          & 80.52          & 43.59          \\ \hline
DeepSeekCoder-FS (6.7B-instruct) & 1.73           & 68.40          & 35.25          & 0.66           & 77.71          & 33.41          & 11.03          & 73.59          & 40.24          \\
Meta-Llama-3.1-FS (8B-instruct)  & 1.32           & 39.11          & 35.34          & 0.58           & 59.93          & 33.33          & 12.14          & 74.28          & 40.37          \\
ChatGPT-FS (gpt-3.5-turbo)       & 1.90           & 73.40          & 37.29          & 1.10           & 86.19          & 36.56          & 13.52          & 77.30          & 42.51          \\ \hline
DivoT5-small (60M)               & 22.31          & 77.29$^\alpha$          & 47.38          & 14.62          & 88.53$^\alpha$          & 42.37          & 17.00          & \textbf{82.68}$^\alpha$ & 43.71          \\
DivoT5-base (220M)               & \textbf{23.02} & 77.65$^\alpha$          & \textbf{47.92} & \textbf{16.55} & 88.58$^\alpha$          & \textbf{43.09} & \textbf{20.90} & 82.47$^\alpha$          & \textbf{46.85} \\ \hline
\end{tabular}
}
% \vspace{-0.4cm}
\end{table*}

% \vspace{-0.1cm}
% \begin{tcolorbox}[left=0cm, right=0cm, top=0cm, bottom=0cm]
% \textbf{Answer to RQ1}: 
\finding{
We conduct experiments on seven datasets for two code editing scenarios. The experimental results demonstrate that the DivoT5-base model significantly outperforms existing models. For example, on the CodeReview dataset, the DivoT5-base model achieves an EM metric improvement of around 10 percentage points compared to the CodeT5-base model. This highlights the superior performance of DivoT5 in capturing code editing characteristics.
}
% \end{tcolorbox}

% \vspace{-0.2cm}

% \vspace{-0.2cm}
\subsection{RQ2: Generalizability to Non-Editing Scenario}
\qy{
To assess the generalization capability of DivoT5 beyond code editing, we introduce the code translation task—a distinct scenario that requires the model to translate code from one programming language to another.
}
In this task, we evaluate the performance of translating Java code to CSharp code and vice versa.
%%% 加具体指数字
Table~\ref{translation} demonstrates that even in the code translation task outside the context of editing existing code, DivoT5-base still exhibits improvements over the existing baseline models.
For example, in the task of translating CSharp code to Java code, DivoT5-base outperforms the CCT5 model by 6.9 percentage points in EM metric, surpasses the CodeReviewer model by 3.2 percentage points, and exceeds the CodeT5-base model by 4.5 percentage points. 
Similarly, in the task of translating Java code to CSharp code, DivoT5-base outperforms CodeT5+~(220M) by over 5\% in the EM metric.
By analyzing the generated results, we find that the translation in this dataset is not simply a direct translation of the original code but also requires corresponding modifications, such as adding the `override` keyword at specific positions, importing and calling packages for certain functions, or even splitting a single line of code. 
\lqy{
Besides, regarding our translation task, the high similarity between Java and CSharp makes the transformation resemble code evolution, where the model decides between copying tokens or generating new ones without drastically altering the structure.
}
These modification patterns are embedded in the training set and are difficult to capture through a few-shot approach, leading to poor performance from billion-scale models. However, when considering translation tasks in specific scenarios, DivoT5 demonstrates an advantage in capturing these modification and translation patterns, performing better than other models. 

\lqy{
Therefore, we believe that these results validate the generalizability of DivoT5 in code-related tasks.
They also highlight how DivoT5 effectively captures both the characteristics of code evolution and the fundamental properties of code itself.
}

% These results validate the generalizability of DivoT5 in code-related tasks. 
% Note that code translation intrinsically differs from code editing tasks, because there is typically no need to perform code translation during the evolution history of a software project. 
% \lqy{
% Note that code translation involves mapping one programming language to another, requiring the model to have a deep understanding of at least one language to produce accurate and meaningful translations.
% }
% Therefore, we believe that DivoT5 not only helps learn the characteristics of code evolution but also the characteristics of code itself.

\begin{table*}[t]
\centering
\renewcommand{\arraystretch}{1.1}
\caption{Results of the code translation task.}
% \vspace{-0.3cm}
\label{translation}
\scalebox{1}{
\begin{tabular}{lcccccc}
\hline
\textbf{}                        & \multicolumn{6}{c}{\textbf{Code Translation}}                                                                              \\ \hline
\textbf{Models}                  & \multicolumn{3}{c}{\textbf{Java to CSharp}}                            & \multicolumn{3}{c}{\textbf{CSharp to Java}}       \\ \hline
\textbf{Metrics}                 & \textbf{EM}    & \textbf{BLEU}  & \multicolumn{1}{c|}{\textbf{C-BLEU}} & \textbf{EM}    & \textbf{BLEU}  & \textbf{C-BLEU} \\ \hline
CodeT5-small (60M)               & 65.40          & 88.23          & \multicolumn{1}{c|}{88.32}           & 69.60          & 87.22          & 87.18           \\
CodeT5-base (220M)               & 66.90          & 88.55          & \multicolumn{1}{c|}{88.72}           & 68.70          & 87.03          & 86.71           \\
CodeT5-large (770M)              & 67.20          & 88.89          & \multicolumn{1}{c|}{88.98}           & 68.80          & 87.20          & 87.16           \\
CodeT5+ (220M)                   & 66.20          & 91.66          & \multicolumn{1}{c|}{91.51}           & 70.20          & 89.64          & 91.01           \\
CoditT5 (220M)                   & 68.30          & 91.45          & \multicolumn{1}{c|}{91.75}           & 70.30          & 90.28          & 90.26           \\
CCT5 (220M)                      & 66.90          & 91.69          & \multicolumn{1}{c|}{91.47}           & 66.30          & 89.40          & 89.21           \\
CodeReviewer (220M)              & 67.80          & 92.06$^\alpha$          & \multicolumn{1}{c|}{91.74}           & 70.00          & 90.22          & 90.22           \\
CodeLlama (6.7B-base)            & 68.40          & 91.53          & \multicolumn{1}{c|}{91.78}           & 68.10          & 90.11          & 87.18           \\ \hline
DeepSeekCoder-FS (6.7B-instruct) & 7.40           & 60.17          & \multicolumn{1}{c|}{57.16}           & 10.30          & 59.51          & 67.68           \\
Meta-Llama-3.1-FS (8B-instruct)  & 6.30           & 59.84          & \multicolumn{1}{c|}{56.98}           & 14.60          & 57.11          & 67.78           \\
ChatGPT-FS (gpt-3.5-turbo)       & 12.80          & 62.53          & \multicolumn{1}{c|}{59.30}           & 18.90          & 62.42          & 72.30           \\ \hline
DivoT5-small (60M)               & 68.30          & 92.03          & \multicolumn{1}{c|}{91.90}           & 72.50          & 91.36$^\alpha$          & 91.38           \\
DivoT5-base (220M)               & \textbf{69.60} & \textbf{92.34}$^\alpha$ & \multicolumn{1}{c|}{\textbf{92.13}}  & \textbf{73.20} & \textbf{91.38}$^\alpha$ & \textbf{91.56}  \\ \hline
\end{tabular}
}
% \vspace{-0.3cm}
\end{table*}

% \vspace{-0.1cm}
% \begin{tcolorbox}[left=0cm, right=0cm, top=0cm, bottom=0cm]
% \textbf{Answer to RQ2}:
\finding{
To demonstrate the generalization capability of DivoT5, we conduct experiments on the code translation task. The results show that DivoT5 significantly outperforms other models in this task.
For example, in translating Java code to CSharp code, DivoT5-base improves the EM score by over 3 percentage points compared to the CodeT5+ model with the same scale.
}
% \end{tcolorbox}

% \vspace{-0.1cm}

\begin{table*}[t]
\centering
\renewcommand{\arraystretch}{1.2}
\caption{Ablation study with DivoT5-small.}
% \vspace{-0.3cm}
\label{ablation}
\scalebox{0.8}{
\begin{tabular}{lccccccccc}
\hline
\textbf{Models}         & \textbf{\begin{tabular}[c]{@{}c@{}}CodeReview\\ EM\end{tabular}} & \textbf{\begin{tabular}[c]{@{}c@{}}NL-CodeRefinement\\ EM\end{tabular}} & \textbf{\begin{tabular}[c]{@{}c@{}}BugFixing-S\\ EM\end{tabular}} & \textbf{\begin{tabular}[c]{@{}c@{}}BugFixing-M\\ EM\end{tabular}} & \textbf{\begin{tabular}[c]{@{}c@{}}Refine-S\\ EM\end{tabular}} & \textbf{\begin{tabular}[c]{@{}c@{}}Refine-S\\ EM\end{tabular}} & \textbf{\begin{tabular}[c]{@{}c@{}}CodeReview-CodeOnly\\ EM\end{tabular}} & \textbf{\begin{tabular}[c]{@{}c@{}}Java-CSharp\\ EM\end{tabular}} & \textbf{\begin{tabular}[c]{@{}c@{}}CSharp-Java\\ EM\end{tabular}} \\ \hline
DivoT5-small$\sim$(60M) & \textbf{40.16}                                                   & \textbf{29.42}                                                          & \textbf{35.84}                                                    & \textbf{28.72}                                                    & \textbf{22.31}                                                 & \textbf{14.62}                                                 & \textbf{17.00}                                                            & \textbf{68.30}                                                    & \textbf{72.50}                                                    \\ \hline
w/o $KSM_{ED}$          & 35.45                                                            & 28.69                                                                   & 35.28                                                             & 27.36                                                             & 21.75                                                          & 13.67                                                          & 14.44                                                                     & 68.00                                                             & 71.20                                                             \\
w/o $RM_{ED}$           & 37.02                                                            & 28.61                                                                   & 35.30                                                             & 28.10                                                             & 21.54                                                          & 13.91                                                          & 13.68                                                                     & 66.30                                                             & 69.80                                                             \\
w/o $EDAE_{ED}$         & 35.22                                                            & 28.74                                                                   & 35.50                                                             & 27.65                                                             & 20.99                                                          & 13.86                                                          & 11.76                                                                     & 67.50                                                             & 68.60                                                             \\
w/o $EDR$               & 37.31                                                            & 28.72                                                                   & 34.76                                                             & 28.37                                                             & 21.68                                                          & 14.03                                                          & 15.83                                                                     & 67.50                                                             & 70.20                                                             \\ \hline
w/o $ALL$               & 34.75                                                            & 28.50                                                                   & 35.35                                                             & 27.36                                                             & 20.84                                                          & 13.41                                                          & 12.05                                                                     & 67.60                                                             & 68.20                                                             \\ \hline
\end{tabular}
}
% \vspace{-0.5cm}
\end{table*}

% Effectiveness of Pre-training task
% \vspace{-0.25cm}
\subsection{RQ3: Effectiveness of Pre-training Tasks}
To evaluate the effectiveness of the proposed pre-training tasks, we conduct an extensive ablation experiment to explore the contribution of each pre-training task. 
Specifically, we remove the pre-training task separately to explore the model's performance. 
Then, we remove all four pre-training tasks to assess the overall impact of pre-training datasets on the model's performance. 
In this setting, our pre-training is to predict the new code on top of the old code.
Due to computational resource limitations, we conduct this ablation study on only the small version of DivoT5.
\qy{
Due to resource constraints, these ablation studies are conducted on the 60M-parameter variant of DivoT5, which provides an efficient setting for experimentation under limited computational resources. This choice is consistent with prior work such as DeepSeekCoder~\cite{guo2024deepseekcoder}, where Fill-in-the-Middle (FIM) ablation experiments were conducted on the 1.3B model—nearly 30× smaller than their largest 33B variant—demonstrating a practical strategy for resource-efficient experimentation.
Furthermore, our approach follows established practices in the CodeT5 model family~\cite{codet52021,zhu2024grammart5}, where 60M-parameter variants are frequently used to examine the effects of different pre-training strategies. Since DivoT5 is also built on the CodeT5 architecture, employing a 60M model for ablation ensures both computational tractability and methodological consistency with previous studies.
% For space limit, we selected three datasets for the ablation studies. All results can be found in our package~\cite{divot5}.
}

\qy{
Table~\ref{ablation} presents the model performance after removing each pre-training task. 
It can be observed that the performance of the model declines when any task is removed, indicating that each pre-training task has a positive influence on the overall effects of the model.
In the NL-guided code editing scenario, removing the $KSM_{ED}$ task (i.e., w/o $KSM_{ED}$) results in a notable decline in EM on the CodeReview dataset—from 40.16\% to 35.45\%. Similarly, removing the $EDR$ task reduces EM to 37.31\%. Comparable trends are observed in other NL-guided datasets such as NL-CodeRefinement and BugFixing, underscoring the general utility of each task in guiding precise edits.
In the code-only editing and non-editing scenarios, similar degradation is observed. For example, on the Refine-S dataset (code-only editing), removing $DAE_{ED}$ lowers the EM from 22.31\% to 20.99\%. On the CSharp$\rightarrow$Java translation task (non-editing), excluding the $RM_{ED}$ task reduces EM from 72.50\% to 69.80\%. These results further support the versatility and necessity of each pre-training task across diverse application settings.
It is worth noting that, compared to removing other pre-training tasks, the removal of $KSM_{ED}$ task has a significant impact on the CodeReview dataset and has the least impact on the code translation task. 
This is reasonable, as the $KSM_{ED}$ task focuses on fine-grained code editing, while the code translation task is distant from the editing scenario, thus suffering from the least impact.
These results indicate that each pre-training task impacts the final performance.
}

Moreover, by comparing the removal of all pre-training tasks~(i.e., w/o $ALL$), we can observe the impact of data on model performance. 
Specifically, we train the w/o $ALL$ version of DivoT5 by directly using the old code as input and the new code as output.
The performance of the w/o $ALL$ model generally falls between that of CodeT5-small and CodeT5-base, indicating that leveraging real code editing history can enhance the model's ability to fulfill code-editing-related tasks to some extent. More importantly, the performance of the w/o $ALL$ model is significantly lower than any ablation version of DivoT5. This demonstrates that our designed directional diffusion strategies within it are crucial for improving model performance than the data alone.

% \vspace{-0.1cm}
% \begin{tcolorbox}[left=0cm, right=0cm, top=0cm, bottom=0cm]
% \textbf{Answer to RQ3}:
\finding{
We compare the effect of removing a specific pre-training task from the DivoT5-small model on the EM metric. We find that the performance of the model decreases to varying degrees on three representative downstream datasets.
This indicates that each pre-training task contributes to the final experimental performance.
}
% \end{tcolorbox}

% \finding{
% We compare the effect of removing a specific pre-training task from the DivoT5-small model on the EM metric. We find that the performance of the model decreases to varying degrees on three representative downstream datasets. This indicates that each pre-training task contributes to the final experimental performance.
% }

% \vspace{-0.25cm}
% \vspace{-0.3cm}

\section{Discussion}\label{discussion}
% compare to diffusion
\lqy{
\textbf{Cost of the Directional Diffusion-Style Model.}
DivoT5, inspired by the gradual denoising concept in diffusion, treats old code during the evolution process as a form of noise and considers the new code as the final noise-free state to simulate real-world code editing scenarios. 
Instead of traditional diffusion, DivoT5 adopts an autoregressive approach, making its training and inference costs the same as encoder-decoder text generation models like Transformer and CodeT5~\cite{transformer2017,codet52021}.
To analyze the training and inference costs, let the sequence length be \(L\)~(i.e., $L_{\text{enc}}$ indicates the input length of the encoder and $L_{\text{dec}}$ indicates the output length of the decoder), the model's hidden layer size be \(H\), and the number of denoising steps in traditional diffusion is \(T\). 
Typically, \(L_{\text{enc}}\) and \(L_{\text{dec}}\) are on the same order of magnitude, so both can be approximated as \(L\).
Thus, the training and inference costs of traditional diffusion are \(O(TL^2H)\), where \(L^2\) reflects interactions between \(L_{\text{enc}}\) and \(L_{\text{dec}}\) in the cross-attention mechanism.
% The training and inference costs of traditional diffusion are \(O(TL^2H)\), where \(L^2\) arises from the global attention mechanism, which typically accounts for interactions between \(L_{\text{enc}}\) and \(L_{\text{dec}}\) due to the cross-attention mechanism.
% The training and inference costs of traditional diffusion are given as \(O(TL^2H)\). Here, \(L^2\) arises from the global attention mechanism, which scales quadratically with the sequence length involved. In an encoder-only model, \(L\) corresponds to \(L_{\text{enc}}\); in a decoder-only model, it corresponds to \(L_{\text{dec}}\). For encoder-decoder models, \(L^2\) typically accounts for interactions between \(L_{\text{enc}}\) and \(L_{\text{dec}}\) due to the cross-attention mechanism.
For an encoder-decoder text generation model with the same structure, the training phase has a time complexity of \(O(L_{\text{enc}}^2H)\)  for the encoder, \(O(L_{\text{dec}}^2H)\) for the decoder, and \(O(L_{\text{enc}}L_{\text{dec}}H)\) for the cross-attention layer. 
% \(L_{\text{enc}}\)/\(L_{\text{dec}}\) denotes the length of the sequence in the encoder/decoder. 
These result in an complexity of \(O(L_{\text{enc}}^2H + L_{\text{enc}}L_{\text{dec}}H + L_{\text{dec}}^2H)\), yielding an overall time complexity of \(O(L^2H)\). In the inference phase, due to the gradual accumulation of the output length, the time complexity increases to \(O(L^3H)\).}

\qy{Above all, the time complexity of the encoder-decoder model during the training phase~(\(O(L^2H)\)) is lower than that of traditional diffusion (\(O(TL^2H)\)). In the inference phase, \(T\) is typically much larger than the sequence length \(L\). For example, to generate text within 200 tokens (\(L\)), \(T\) may need to be set to 2000~\cite{difflm_ARDiff_wu2024ar, difflm_DiffLM_li2022diffusion}. Consequently, \(O(L^3H)\) is often significantly smaller than \(O(TL^2H)\). In summary, the training and inference costs of DivoT5 are generally lower than those of traditional diffusion, and can therefore also be effectively adopted for scaling to larger models or alternative architectures.
}

% data impact
\lqy{
\textbf{Impact of Data and Pre-training Tasks.}
To better reflect real-world editing scenarios, we incorporate code data related to actual changes, which may directly enhance model performance. 
To isolate the impact of this data, we removed all designed pretraining tasks in an ablation study. As shown in Table~\ref{ablation}, while the new data helped improve performance~(EM score increased from 32.25 to 34.75 on the CodeReview dataset), the substantial boost from our diffusion strategy~(EM score of 40.16) highlights that the design of DivoT5 is more crucial than the dataset. 
Therefore, we believe that DivoT5's performance is primarily driven by leveraging the diffusion style to simulate real editing scenarios, rather than by the data itself.
}

\minorqy{
\textbf{Generalization and Extensibility.}
DivoT5 is designed with strong generalization potential across both code granularity levels and model architectures. While it is trained primarily on token-level code edits, this level of granularity serves as the foundation for modeling more complex transformations. 
In real-world software maintenance, many structural changes, such as line-level updates, block replacements, or module-level refactorings, are composed of sequences of fine-grained edits. By modeling token-level operations, DivoT5 captures the fundamental building blocks of such transformations and has potential extensions toward hierarchical or semantic-aware edit modeling. Recent studies~\cite{edit_emperical,edit_emperical_2} also emphasize the prevalence and practical significance of token-level modifications in real development workflows, further validating our design choice.
In addition, DivoT5 aims to simulate real-world developer behavior by leveraging the intermediate states of code evolution as training signals. Instead of relying solely on before-and-after code pairs, we extract supervision from the actual editing trajectory, which more closely reflects how developers iteratively modify code. These evolution directions are architecture-agnostic and can be applied to various model types. Although we use an encoder-decoder architecture in this work, the same training data and objectives can be easily adapted to other architectures~(e.g., decoder-only architecture) or larger language models. This design makes our approach flexible, extensible, and compatible with future model developments.
}

\section{Threats to Validity}

% \textbf{Threats to internal validity} might comes from four factors.
% 基础模型的选择
% 预训练数据质量
% 预训练的实现问题——performance
% 数据泄露问题
% The first factor is the selection of the base model, which can impact the final experimental results. To mitigate this potential threat, we choose popular and representative model, CodeT5, as the base model. CodeT5 is widely recognized as a classic encoder-decoder framework and has been extensively applied in various tasks\cite{codet52021}. The second factor is the quality of the pre-training data, as low-quality code changes can potentially mislead the model's learning process. To address this concern, we select well-maintained open-source projects and implement data cleaning strategies to mitigate this threat. The third factor pertains to the implementation of our experiments. To mitigate this threat, we adopt the same evaluation metrics as the original paper to assess the model's performance. Additionally, we fix the random seed to ensure experiment reproducibility. The final factor is related to data leakage. To mitigate this threat, we carefully select a small number of popular open-source projects. These projects have already been used in other pre-training datasets, ensuring that they do not introduce additional data leakage concerns.

\textbf{Threats to internal validity} might come from the implementation of our experiments. 
% To mitigate this threat, we rely on the the same evaluation metrics reported in the original paper to compare the performance of different models. 
To mitigate this threat, we utilize open-source models available on HuggingFace~\cite{huggingface} to minimize potential errors.
Additionally, we fix the random seed to ensure experiment reproducibility.

\textbf{Threats to external validity} may arise from two factors.
The first factor is the selection of the base model. To mitigate this potential threat, we choose a popular and representative model, CodeT5, as the base model. CodeT5 is widely recognized as a classic encoder-decoder framework and has been extensively applied in various tasks~\cite{codet52021}. 
\qy{
The second factor lies in the downstream datasets used in our experiment. To mitigate this threat, we choose the widely used datasets from CodeXGLEU~\cite{codexglue2021} and other existing code evolution models~\cite{coditt52022, codereviewer}. To further generalize our results, we need to evaluate DivoT5 on more datasets. 
Nonetheless, we acknowledge that important code editing tasks, such as code refactoring, which improves structure without changing functionality, are not covered in this work. In future work, we plan to scale DivoT5 to larger models and extend it to broader scenarios, including refactoring and architectural changes.
As DivoT5 is designed for gradual code evolution with both copying and generation, its advantages are less evident in tasks with large input-output gaps, such as natural language to code or vice versa.
}
% Since DivoT5 focuses on the gradual evolution from code to code, considering both copying and generating tokens, its performance improvement is minimal on tasks involving the transformation between massively different input and output, such as natural language to code or vice versa. 

% The third factor is the quality of the pre-training data, as low-quality code changes can potentially mislead the model's learning process. To address this concern, we select well-maintained open-source projects and implement data cleaning strategies to mitigate this threat. The third factor pertains to the implementation of our experiments.
% The third factor is the choice of programming language for pre-training data. To address this concern, we select Java, one of the most widely used programming languages, to collect the pre-training data. As the Java programming language is prevalent in the benchmarks of most code editing downstream tasks, we choose Java as the target programming language for our experiments. We may need additional data collection and more experiments for generalization of our results to other programming languages.

\textbf{Threat to construct validity} may arise from two factors. 
The first is the selection of evaluation metrics. To mitigate this concern, we choose widely used evaluation metrics in the field of code generation to assess the evolved code. For the CodeReview dataset, we also select the same evaluation metrics as other code evolution models~\cite{coditt52022}.
The second is related to data leakage. To mitigate this threat, we exclude data that overlaps with downstream tasks during the pre-training process. \lqy{
Specifically, we perform precise deduplication using a string-matching approach. If a sample from the downstream task’s test set appears in any pretraining data sample `x', we filter `x' out. 
}
Additionally, we conduct extensive ablation experiments to validate the effectiveness of our technical contributions.

% 下游任务数据集的代表性
% 模型的可迁移性
% 评估指标的选择
% The first factor lies in downstream datasets used in our experiment. To mitigate this threat, we choose the widely used datasets from CodeXGLEU~\cite{codexglue2021} and other existing code evolution models~\cite{coditt52022}. 
% The second factor is the selection of evaluation metrics. To mitigate this concern, we choose widely used evaluation metrics in the field of code generation to assess the evolved code. For the automated code review task, we also select the same evaluation metrics as other code evolution models~\cite{coditt52022}.
% The third factor is that currently, DivoT5 is only evaluated on the selected tasks and datasets. Whether it can generalize to other tasks in the field of SE is still unknown. 
% However, DivoT5 demonstrate significant improvements over the CodeT5 model and other code editing-related models on multiple downstream tasks. This indicates that incorporating the basic editing intent can effectively enhance the model's ability to understand code evolution.

% \vspace{-0.25cm}
\section{Conclusion}
In this paper, we propose DivoT5, a directional diffusion-style pre-trained model designed to simulate the incremental nature of human code editing and effectively capture the characteristics of code evolution.
% Our DivoT5 captures evolutionary characteristics of code editing. 
Specifically, we first propose three pre-training tasks to make the model aware of fine-grained code editing information (i.e., the direction of code evolution) and enrich the context of the old code. 
Then, we design the fourth pre-training task utilizing intermediate states during the editing process to further reinforce the direction of code evolution.
% We first propo is the dataset, we collect high-quality code editing datasets, JCommitPatch, from real-world evolution scenarios. These data cover the basic patterns of code changes and the basic intentions of programmers to modify code.
% The second aspect is pre-training objectives. We design three fine-grained edit-aware pre-training objectives to capture the underlying editing intentions contained in the data. 
% Each objective serves to train the model in understanding and capturing different aspects of code editing intentions.
Experiments are conducted on two code-editing scenarios and one non-editing scenario using eight datasets, 
demonstrating that DivoT5 achieves state-of-the-art performance on most tasks compared to fine-tuned similar or larger scale models~(60M, 220M, 770M, 6.7B),  and few-shot billion-scale instruct models~(6.7B, 8B, ChatGPT).
In addition, our ablation experiments also confirm the contribution of each pre-training task to the final performance.

\section{Acknowledgment}
This work is sponsored by the National Key Research and Development Program of China under Grant No. 2023YFB4503803, the National Natural Science Foundation of China under Grant No. 62232003 and No. 62402482, and the CCF-Huawei Populus Grove Fund (Grant No. CCF-HuaweiSE202407).

% \section{Data Availability}
% We make DivoT5 publicly available on website~\url{https://github.com/LIANGQINGYUAN/DivoT5}

\bibliographystyle{IEEEtran}
\bibliography{ref.bib}

\end{document}